%% file: main.tex
\documentclass{article}
\usepackage{ijcai22}

\usepackage{times}

\usepackage{soul}
\usepackage{url}
\usepackage[hidelinks]{hyperref}
\usepackage[utf8]{inputenc}
\usepackage[small]{caption}
\usepackage{graphicx}
\usepackage{amsmath}
\usepackage{booktabs}
\input{utils}

\title{The Dichotomous Affiliate Stable Matching Problem: Approval-Based Matching with Applicant-Employer Relations}

\author{
Marina Knittel\and Samuel Dooley\and John P. Dickerson
\affiliations
Computer Science Department, University of Maryland
\emails
\{mknittel, sdooley1, john\}@cs.umd.edu
}

\begin{document}

\maketitle
\begin{abstract}
\input{abstract}
\end{abstract}

\input{introduction}

\input{model_defn}

\input{survey}

\input{theory}

\input{experiments}

\input{conclusions}

{
\bibliographystyle{named}
\bibliography{references}
}

\ifappendix
\clearpage
\input{appendix}

\fi

\end{document}

%% file: utils.tex
\usepackage{amsfonts, amsmath, amsthm, enumerate, gensymb, mathtools, tikz, graphicx, thmtools, algorithmic, algorithm, caption, subcaption}
\usepackage{adjustbox}
\usepackage{booktabs}
\usetikzlibrary{positioning,chains,fit,shapes,calc}

\usepackage{setspace}
\usepackage{enumitem}
\setitemize{noitemsep,topsep=0pt,parsep=0pt,partopsep=0pt}

\newtheorem{theorem}{Theorem}

\newtheorem{lem}{Lemma}
\newtheorem{prop}{Proposition} 

\newtheorem{definition}{Definition}

\definecolor{forestgreen}{rgb}{0.13, 0.54, 0.13}

\newcommand{\T}{\mathcal{T}}
\newcommand{\E}{\mathcal{E}}
\newcommand{\I}{\mathbb{I}}

\newcommand{\pr}{\mathsf{pr}}
\newcommand{\aff}{\mathsf{aff}}

\newcommand{\sprm}{\textsf{SmartPriorityMatch}}
\newcommand{\prm}{\textsf{PriorityMatch}}
\newcommand{\agg}{\textsf{val}}

\newcommand{\afdef}{\textsf{val}}
\newcommand{\val}{\textsf{val}}

\newcommand{\bamlong}{Dichotomous Affiliate Stable Matching}
\newcommand{\bam}{\textsc{DASM}}
\newcommand{\bamprob}{\bam{} Problem}

\newif\ifappendix
\appendixtrue

%% file: abstract.tex
While the stable marriage problem and its variants model a vast range of matching markets, they fail to capture complex agent relationships, such as the affiliation of applicants and employers in an interview marketplace. To model this problem, the existing literature on matching with externalities permits agents to provide complete and total rankings over matchings based off of both their own and their affiliates' matches. This complete ordering restriction is unrealistic, and further the model may have an empty core. To address this, we introduce the Dichotomous Affiliate Stable Matching (\bam{}) Problem, where agents' preferences indicate dichotomous acceptance or rejection of another agent in the marketplace, both for themselves and their affiliates.  
We also assume the agent's preferences over entire matchings are determined by a general weighted valuation function of their (and their affiliates') matches. Our results are threefold: (1) we use a human study to show that real-world matching rankings follow our assumed valuation function; (2) we prove that there always exists a stable solution by providing an efficient, easily-implementable algorithm that finds such a solution; and (3) we experimentally validate the efficiency of our algorithm versus a linear-programming-based approach.

%% file: introduction.tex
\section{Introduction}\label{sec:intro}
In many markets, two classes of participants seek to be paired with each other.  For example, in labor markets, workers are paired with firms~\cite{Perrault16:Strategy}; in online advertising, eyeballs are paired with advertisements~\cite{Shen20:Learning,Dickerson19:Online}; and, in morally-laden settings such as refugee resettlement and organ donation, refugees are paired with new housing locations~\cite{Jones18:Local} and donors are paired with needy recipients~\cite{Ashlagi21:Kidney,Li14:Egalitarian}, respectively.  The field of market design purports to provide analytically-sound and empirically-validated approaches to the design and fielding of such matching markets, and necessarily joins fields such as economics and computer science~\cite{Roth02:Economist,Roth18:Marketplaces}.

The seminal work of Gale and Shapley~\shortcite{Gale62:College} characterized the stable marriage problem, where both sides of a market---workers and firms, refugees and settlement locations, etc.---express preferences over the other side, and the goal is to find a robust matching that does not unravel in the face of agents' selfish behavior.  Myriad generalizations were proposed in the following decades;  see Manlove~\shortcite{Manlove13:Algorithmics} for an overview of the history and variants of these problems.  Largely, these models assume that agents' preferences only consider the direct impact of an outcome on that agent.

One extension of stable marriage is matching with externalities wherein agents on each side of a two-sided market have preferences over their own match \emph{and} the matches of others. These models often incorporate many more realistic and complex assumptions which makes for a richer and harder to analyze matching setting~\cite{pycia2012stability,echenique2007solution,baccara2012field}. Sasaki and Toda~\shortcite{sasaki1996two} first introduced matching with externalities, where agents' decisions to deviate from a proposed match depended on reasonable assumptions for the reaction of other agents to the deviation. Hafalir~\shortcite{hafalir2008stability} and Mumcu and Saglam~\shortcite{mumcu2010stable} expand upon this stability notion for one-to-one matchings with further restrictions on agent behavior; while Bando~\shortcite{bando2012many,bando2014modified} extends the analysis to many-to-one matchings where firms consider other firms' externalities.

Much work in the matching with externalities literature focuses on the appropriateness of various stability definitions. Much analysis then centers on the complexity and hardness of the proposed matching algorithms. For instance, Br\^{a}nzei~\shortcite{branzei2013matchings} models agent values in matching with externalities as arbitrary functions and creates a valuation as a sum over the agent's values over all matches. In our work, an agent values a match as either acceptable or unacceptable (dichotomously), and we do a (weighted) sum over all relevant matches for the agent to get their valuation.
While there is existing work on the complexities of these matchings, eliciting general preferences over a complex market can be intractable, both with respect to human ability and computational/communication complexity~\cite{Rastegari16:Preference,Sandholm06:Preference}. One commonly imposed assumption is that of dichotomous preferences \cite{Bogomolnaia04:Random}, which coarsely places alternatives into acceptable or unacceptable bins. 

This work is inspired by Dooley and Dickerson~\shortcite{Dooley20:Affiliate}, which explores matching with externalities in academic faculty hiring; however, in our work, we analyze the marketplace with dichotomous preferences. Our main motivation is the academic faculty \textit{interview} marketplace, where we match interview slots for universities and graduating students, and universities care about their graduating students' matches. Other motivations include playdate matching, study abroad, student project allocation, and the dog breeding market.

We note that the only simplification we introduce to the Dooley and Dickerson model is that of binary preferences. This assumption is prevalent in various matching settings like resource allocation~\cite{Ortega20:Multi-unit} and more specifically in the allocation of unused classrooms in a school setting~\cite{Kurokawa18:Leximin} and barter exchange~\cite{Aziz20:Strategyproof}. With this additional assumption, we are able to provide positive and constructive principled approaches to clearing (dichotomous) affiliate matching markets. 

\noindent\textbf{Our contributions.}  We view our contributions as follows.
\begin{itemize}
    \item We introduce the \bamlong{} (\bam{}) Problem, which characterizes the affiliate matching problem under dichotomous preferences to better accommodate realistic preference elicitation constraints, along with a valuation function for agents to rank matches based off their preferences, parameterized by an employer's relative valuation of its affiliates' and its own matches (\S\ref{sec:model});
    \item We run a human survey to provide support for the model design choices, showing that real people in some situations may, indeed, adhere to our valuation function under different parameters (\S\ref{sec:human-experiments});
    \item We propose an efficient algorithm to solve the \bamprob{} (\S\ref{sec:theory}), i.e., yield a stable matching; and
    \item We perform experimental validation of our algorithmic approach to verify its correctness and scalability  (\S\ref{sec:experiments}).
\end{itemize}

%% file: model_defn.tex
\section{Model Definition}\label{sec:model}
We now present our matching model.  
This model represents hiring interview markets where applicants have previous affiliations with employers and preferences are encoded as binary values that denote interest or disinterest (i.e., \emph{dichotomous preferences}~\cite{Bogomolnaia04:Random}). We describe the model in the most general many-to-many setting. We formalize the model, including defining valuation functions (\S\ref{sec:model-prelims}), stability, and other useful concepts (\S\ref{sec:model-defs}).

\subsection{The \bamlong{} Problem}\label{sec:model-prelims}
In the \bamlong{} (\bam{}) Problem, we are given sets $A$ of $n$ applicants and $E$ of $m$ employers. For every $a\in A$ (resp. $e\in E$), we are given a complete preference function $\pr_a: E \to \{0,1\}$ (resp.\ $\pr_e^e:A\to \{0,1\}$, the notational difference will be clear later). If $u$ and $v$ are on opposite sides, then we say $u$ is \textit{interested in} or \textit{likes} $v$ if $\pr_u(v)=1$ (or $\pr_u^u(v)=1$ if $u\in E$), otherwise $u$ is \textit{disinterested in} $v$. The \textit{many-to-many} matching scenario specifies that $u$ might be matched with as many as $q(u)$ agents on the other side of the market, where $q(u)$ is the \textit{capacity} of $u$. A valid matching is a function $\mu:A\cup E \to 2^{A\cup E}$ such that for any $a\in A$ (resp. $e\in E$): $\mu(a) \subseteq E$ (resp. $\mu(e)\subseteq A$), $|\mu(a)| \leq q(a)$ (resp. $|\mu(e)| \leq q(e)$), and $e\in \mu(a)$ if and only if $a\in \mu(e)$. %

One defining aspect of this market is the notion of \textit{affiliates} which represent previous relationships between agents. Let $\aff: E\to 2^A$ return an employer's set of affiliate. For instance, in Figure~\ref{fig:example}, $a_1$ is $e_1$'s affiliate, and $a_2$ and $a_3$ are both $e_2$'s affiliates. Then $\aff(e_1) = \{a_1\}$, $\aff(e_2) = \{a_2,a_3\}$, and $\aff(e_3)=\emptyset$. Note that $\aff$ over all $e\in E$ forms a disjoint cover of $A$, so each applicant is the affiliate of exactly one employer. In this model, $e$ cares about its affiliates' matches. To express this, for any $a\in \aff(e)$, $e$ has preferences $\pr_e^a: E\to\{0,1\}$. To account for these preferences, $e$'s valuation of matchings is over tuples of its own \textit{and} its affiliates' matches. While these valuations may be general, we will examine a natural and flexible additive valuation method.

\begin{definition}
For any $e\in E$ and $a\in A$, we define the \textbf{weighted valuation function} over a match $\mu$ for a given weight $\lambda \in [0,1]$ as:
\begin{align*}
\agg_e(\mu) =& \sum_{a^*\in \mu(e)} \pr_e^e(a^*) + \lambda\sum_{a_i\in\aff(e)} \sum_{e^*\in\mu(a_i)} \pr_e^{a_i}(e^*)\\
 \val_a(\mu) =& \sum_{e^*\in\mu(a)} \pr_a(e^*).
\end{align*}
And we say $e$ or $a$ \textbf{prefers} $\mu$ to $\mu'$ ($\mu \succ_{e} \mu'$ or $\mu \succ_a \mu'$) if and only if $\agg_e(\mu) > \agg_e(\mu')$ or $\val_a(\mu) > \val_a(\mu')$.
\end{definition}

\input{example_fig}

This function does not necessarily create a total order over matchings, as an agent may have the same valuation for two distinct matchings. Then it does not prefer one matching to another. In the employer version, $\lambda$ parameterizes how employers weigh the value of their affiliates' matches with respect to their own matches. If $\lambda=1$, then an employer cares about each affiliate match as much as each of its own matches. If $\lambda=0$, employers do not care about their affiliates' matches. Setting $\lambda=\epsilon$ for small $\epsilon>0$ yields a lexicographic valuation where employers care about their matches first and only use affiliates' matches as tiebreakers.

To understand the role of $\lambda$, consider again our example in Figure~\ref{fig:example} and let $\lambda=1$. Then $e_2$'s valuations of $\mu$ and $\mu'$ are: $\agg_{e_2}(\mu) = 3$ (since it likes one of its matches and both of its affiliates' matches) and $\agg_{e_2}(\mu') = 2$ (since it likes both of its matches). Therefore $\mu \succ_{e_2} \mu'$. If $\lambda = \epsilon$, then $e_2$'s valuations of the matchings are: $\afdef_{e_2}(\mu)  = 1 + 2\epsilon$ and $\afdef_{e_2}(\mu') = 2$. This means $\mu \prec_{e_2} \mu'$. As we discuss later, this illustrates how $\lambda$ can affect which matchings are stable.

\subsection{Blocking Tuples and Stability}\label{sec:model-defs}
Next, we define the notion of stability in the \bamprob{}. As in the standard stable marriage problem~\cite{Gale62:College} and its many variants, our notion of stability relies on the (non)existence of \emph{blocking tuples}.  The blocking tuple---traditionally, blocking \emph{pair}---is designed to identify a set of unmatched individuals who might cheat in order to match with each other. We formalize cheating as follows:

\begin{definition}
Consider an instance of the \bamprob{} with matching $\mu$. An agent $a\in A\cup E$ \textbf{cheats} if they break a match with some agent $a'\in \mu(a)$.
\end{definition}

Like in the standard stable marriage problem, we would like to ensure that no two agents\footnote{In future work, this could be generalized to \textit{coalitions}, or larger sets of agents.} $a\in A$ and $e\in E$ will agree to cheat on their assigned match to (possibly) match with each other assuming no other agents cheat. In stable marriage, the only way (up to) two agents will cheat, and thus cause instability, is if they prefer each other to their matches.

In our setting, stability is not as simple. Let $a\in A$ and $e\in E$ be two agents who consider cheating on matches $e'\in \mu(a)$ and $a'\in \mu(e)$ respectively. Once $e'$ and $a'$ lose their matches with $a$ and $e$, they could naturally consider filling that empty space in their capacities by matching with each other (this is not cheating). In stable marriage, we do not need to model $e'$'s and $a'$'s reaction because this does not impact $e$'s and $a$'s decisions to cheat. In the \bamprob{}, however, $e$'s preference profile is more complicated. For instance, if $a'\in \aff(e)$, and $e$ wants $a'$ to be matched with $e'$, $e$ might decide to cheat on $a'$ so that $a'$ will fill its new empty capacity by matching with $e'$. Note that this still does not require $a'$ or $e'$ to cheat, since they are only forming matches instead of breaking them. %

To this end, we define our notion of the blocking tuple to capture not only the cheating between two agents, but also the responses of those whose matches have been broken.

\begin{definition}\label{def:blocktup}
Consider an instance of the \bamprob{} with matching $\mu$ and some tuple $\T=(a_1,\ldots,a_k)$ where $a_i\in A\cup E$ for all $i\in[k]$. Construct $\mu'$ with the following process:
\begin{enumerate}
\item Allow up to two agents in $\T$ to cheat on one match each. If exactly two agents cheat, they then match with each other. Otherwise, the sole cheater may match assuming this new match does not violate any capacities.
\item All other agents in $\T$ are then allowed to form new matches up to their capacity.
\end{enumerate}
Then $\T$ is a \textbf{blocking tuple} if: (1) all cheating agents strictly prefer $\mu'$ to $\mu$, and (2) for any other agent $a\in \T$ that forms a match with $a' \in \T$, $a$ strictly prefers $\mu'$ to $\mu'\setminus\{(a,a')\}$. An instance of the \bamprob{} is \textbf{stable} if and only if it contains no blocking tuples.
\end{definition}

In stable marriage, this becomes the same standard notion of cheating, and therefore this is a natural extension of the stable marriage blocking pair. In Definition~\ref{def:blocktup}, a cheating agent would only cheat if they know other agents could respond in a way such that the resulting matching is preferable to the original matching. After cheating occurs, non-cheating agents will respond with a match if they would prefer to have that match in the final matching. Thus, non-cheating agents are not performing calculated activity; they are simply reacting.

Interestingly, we only need to consider tuples of size at most six that satisfy certain properties to determine if a blocking tuple exists. See Proposition~\ref{def:block} and Appendix~\ref{appendix:model} for reasoning and a proof. These properties are captured by the \textit{potential blocking tuple}, a tuple of size at most six with the appropriate agents such that, given the right preference profiles, they could be blocking tuples. To accommodate the sextuplet notation, we introduce the ``empty agent'', $\gamma$, and a set $\mathcal{E}=\{\gamma\}$, which represents the absence of an agent. Formally, $\gamma$ is an agent in the system with no side or affiliation and zero capacity. Additionally, given a matching $\mu$, we must notate the agents who have remaining capacity. Let $N_\mu^A = \{a\in A: |\mu(a)| < q(a)\}$ and $N_\mu^E=\{e\in E: |\mu(e)| < q(e)\}$.

\begin{definition}\label{def:potential-blocking-tuple}
Consider an instance of the \bamprob{} with matching $\mu$. A tuple $\T = (a,a',a'',e,e',e'')$ is a \textbf{potential blocking tuple} for $\mu$ if all the following hold:

\begin{enumerate}
\item $a\in A$
\item $e\in E\setminus\mu(a)$
\item $a'\in \mu(e)\cup \mathcal{E}$, where $a'\in \mathcal{E}$ only if $e\in N_\mu^E$
\item $e'\in \mu(a) \cup \mathcal{E}$, where $e'\in\mathcal{E}$ only if $a\in N_\mu^A$
\item $a''\in (N_\mu^A\cup\mathcal{E}\cup\{a'\})\setminus\mu(e')$, where $a''\in\mathcal{E}$ if $e'\in\mathcal{E}$
\item $e''\in (N_\mu^E\cup \mathcal{E}\cup\{e'\})\setminus\mu(a')$, where $e''\in\mathcal{E}$ if $a'\in\mathcal{E}$
\item $a''=a'\notin\mathcal{E}$ if and only if $e''=e'\notin\mathcal{E}$
\end{enumerate}

\end{definition}

We now clarify the purpose of each condition respectively:
\begin{enumerate}[itemsep=0mm]
\item $a$ must be an applicant.
\item $e$ must be an employer that is not matched with $a$ (else they cannot form a blocking tuple).
\item $a'$ is $e$'s old match that is broken. If $e$ simply has additional capacity, then $a'\in\mathcal{E}$ is an empty agent.
\item $e'$ is $a$'s old match that is broken. If $a$ simply has additional capacity, then $e'\in\mathcal{E}$ is an empty agent.
\item $a''$ is $e'$'s new match. It must have unmatched capacity, or (if $e'$ and $a'$ decide to match) is instead $a'$ itself. It could be an empty agent if $e'$ does not rematch (and must be if $e'$ is an empty agent). Additionally, we must ensure it was not previously matched to $e'$.
\item $e''$ is $a'$'s new match. It must have unmatched capacity, or (if $e'$ and $a'$ decide to match) is instead $e'$ itself. It could be an empty agent if $a'$ does not rematch (and must be if $a'$ is an empty agent). Additionally, we must ensure it was not previously matched to $a'$.
\item If $e'$ matches with $a'$ (where $a'=a''\notin\mathcal{E}$), then $a'$ must match with $e'$ (where $e'=e''\notin\mathcal{E}$).
\end{enumerate}

Consider Figure \ref{fig:example} and tuple $ (a_3,a_2,a_2,e_2,e_1,e_1)$. Some agents are duplicated in this tuple; this is okay. This potential blocking tuple describes the following changes: (1) $a_3$ breaks its match with $e_1$, (2) $e_2$ breaks its match with $a_2$, (3) $a_3$ and $e_2$ match together, and (4) $a_2$ and $e_1$ match together. Note that the last part occurs because $a_2$ and $e_1$ appear twice in the tuple. If we rewrite the tuple as $ (a_3,a_2,a_2',e_2,e_1,e_1')$ to distinguish duplicate instances, then $a_2'$ indicates that $e_1$ matches with $a_2$ and $e_1'$ indicates $a_2$ matches with $e_1$.

Definition~\ref{def:potential-blocking-tuple}'s matching constraints ensure that broken and formed matches in this process make sense (e.g., no two agents will be matched to each other twice). When we consider a blocking tuple, we must compare the matching to the alternative matching that occurs after swapping as described.

\begin{definition}
Consider an instance of the \bamprob{} with matching $\mu$ and a potential blocking tuple $\mathcal{T} = (a,a',a'',e,e',e'')$. Let $\mu'$ be defined by starting at $\mu$, breaking the matches $(a,e')$ and $(a',e)$, adding match $(a,e)$, and adding matches $(a',e'')$ and $(a'',e')$ if and only if those variables are not in $\mathcal{E}$ respectively. Then $\mu'$ is the \textbf{swapped matching} of $\mu$ with respect to $\T$.
\end{definition}

In Figure~\ref{fig:example}, $\mu'$ is the swapped matching of $\mu$ with respect to $(a_3, a_2, a_2, e_2, e_1, e_1)$. Note that it is a valid matching.

\begin{prop}\label{prop:matching}
If $\mu$ is a matching and $\mu'$ is the swapped matching of $\mu$ with respect to some potential blocking tuple $\mathcal{T}$, then $\mu'$ is a matching.
\end{prop}

See Appendix~\ref{appendix:model} for a proof. Now we show that the set of potential blocking tuples are sufficient consideration to show that an instance of the \bamprob{} is unstable.

\begin{prop}\label{def:block}
Consider an instance of the \bamprob{} with matching $\mu$. Then $\mu$ is unstable if and only if there exists a potential blocking tuple $\mathcal{T} = (a,a',a'',e,e',e'')$ with respective swapped matching $\mu'$ for $\mu$ such that:
\begin{enumerate}
\item $\mu \prec_a \mu'$
\item $\mu \prec_e \mu'$
\item If $a'\notin\mathcal{E}$, then $\mu'\setminus\{(a',e'')\} \prec_{a'} \mu'$
\item If $e'\notin\mathcal{E}$, then $\mu'\setminus\{(a'',e')\} \prec_{e'} \mu'$
\item If $a''\notin\mathcal{E}$, then $\mu'\setminus\{(a'',e')\} \prec_{a''} \mu'$
\item If $e''\notin\mathcal{E}$, then $\mu'\setminus\{(a',e'')\} \prec_{e''} \mu'$
\end{enumerate}
\end{prop}

See Appendix~\ref{appendix:model} for a proof. In the rest of the paper, we assume all blocking tuples take this form. The first two conditions ensure that $a$ and $e$ prefer the new match $\mu'$ to $\mu$. The next four state that in the context of $\mu'$, all of $a',e',a'',$ and $e''$ must actively desire the new match. In these conditions, we only care if $a'$, $a''$, $e'$, and $e''$ are not in $\mathcal{E}$ (i.e., they exist).

Consider again Figure~\ref{fig:example} when $\lambda=1$. Recall that $\T = (a_3, a_2, a_2, e_2, e_1, e_1)$ is a potential blocking tuple for $\mu$ and $\mu'$ is the swapped matching of $\mu$ with respect to $\T$. Under our weighted valuation function with $\lambda=1$, we already showed that $e_2$ prefers $\mu'$ to $\mu$. Additionally, since $a_3$ doesn't like its match in $\mu$ but likes its match in $\mu'$, it also prefers $\mu'$. Once $a_3$ and $e_2$ have broken their matches with $a_2$ and $e_1$ respectively, $a_2$ and $e_1$ have an active interest in matching. This implies that $\T$ satisfies all constraints in Proposition~\ref{def:block} and is a blocking tuple. When $\lambda=\epsilon$, since $e_2$ does not prefer $\mu'$ to $\mu$, this is not a blocking tuple.

%% file: example_fig.tex
\definecolor{myblue}{RGB}{80,80,160}
\definecolor{myred}{RGB}{160,80,80}

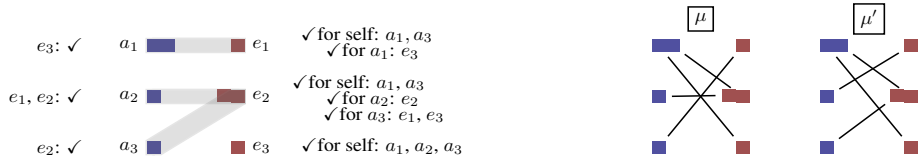
\begin{figure*}
    \centering
    \resizebox{35em}{6em}{
    \begin{tikzpicture}[thick,
      asnode/.style={fill=myblue},
      esnode/.style={fill=myred},
      every fit/.style={ellipse,draw,inner sep=-2pt,text width=2cm},shorten >= 3pt,shorten <= 3pt
    ]
    
    \begin{scope}[start chain=going below,node distance=7mm]
    \node[asnode,on chain] (a1) [label=left: $a_1$] {};
    \node[asnode] (a11) at (.25,0)  {};
    \foreach \i in {2,3}
      \node[asnode,on chain] (a\i) [label=left: $a_\i$] {};
    \end{scope}
    \node[left= of a1] { $e_3$: \checkmark};
    \node[left= of a2] { $e_1$, $e_2$: \checkmark};
    \node[left= of a3] { $e_2$: \checkmark};
    
    \begin{scope}[xshift=1.5cm,start chain=going below,node distance=7mm]
    \node[esnode,on chain] (e1) [label=right: $e_1$] {};
    \node[esnode,on chain] (e2) [label=right: $e_2$] {};
    \node[esnode] (e21) at (-.25, -.94) {};
    \node[esnode,on chain] (e3) [label=right: $e_3$] {};
    \end{scope}
    \node (we1) at (3.8,.2){\checkmark for self: $a_1, a_3$};
    \node (we1a) at (3.9,-.1) {\checkmark for $a_1$: $e_3$};
    \node (we2) at (3.65,-.7){\checkmark for self: $a_1$, $a_3$};
    \node (we2a) at (3.9,-1.0) {\checkmark for $a_2$: $e_2$};
    \node (we2b) at (4.1,-1.3) {\checkmark for $a_3$: $e_1$, $e_3$};
    \node at (4.07,-1.9) {\checkmark for self: $a_1$, $a_2$, $a_3$};
    
    \draw[draw=gray, thick, fill=gray!80!black, opacity=0.2]
        (a1.north west) -- (e1.north east) -- (e1.south east) -- (a1.south west) -- (a1.north west) -- (a1.north) -- (e1.north east);
    \draw[draw=gray, thick, fill=gray!80!black, opacity=0.2]
        (a2.north west) -- (e2.north east) -- (e2.south east) -- (a2.south west) -- (a2.north west) -- (a2.north) -- (e2.north east);
    \draw[draw=gray, thick, fill=gray!80!black, opacity=0.2]
        (a3.south west) -- (a3.south east) -- (e2.south east) -- (e2.north east) -- (e2.north west) -- (a3.north west) -- (a3.south west) -- (a3.south east);
        
    \foreach \j in {2,3}
    {%
        \pgfmathsetmacro{\aj}{3*\j+3}
        \pgfmathsetmacro{\ej}{3*\j+4+.5}
        \pgfmathsetmacro{\mup}{3*\j+3.5}
        \begin{scope}[xshift=\aj cm, start chain=going below,node distance=7mm]
        \foreach \i in {0,1,2} 
          {
          \node[asnode,on chain] (a\i\j) {};
          }
        \end{scope}
        \begin{scope}[xshift=\ej cm,start chain=going below,node distance=7mm]
        \foreach \i in {0,1,2}
          {
          \node[esnode,on chain] (e\i\j) {};
          }
        \end{scope}
    }
    \node[asnode] (a02a) at (9.25, 0) {};
    \node[esnode] (e12a) at (10.25, -.94) {};

    \node[draw] at (9.75,0.5) (mu2)  {$\mu$};
    \path[-] (a02a) edge (e12);
    \path[-] (a02) edge (e22);
    \path[-] (a12) edge (e12a);
    \path[-] (a22) edge (e02);

    \node[asnode] (a03a) at (12.25, 0) {};
    \node[esnode] (e13a) at (13.25, -.94) {};
    \node[draw] at (12.75,0.5) (mu2)  {$\mu'$};
    \path[-] (a03a) edge (e13);
    \path[-] (a03) edge (e23);
    \path[-] (a13) edge (e03);
    \path[-] (a23) edge (e13a);

    \end{tikzpicture}   
    }
    \caption{An example matching problem. On the left are affiliations and preferences for each agent. The capacity of each agent is represented with squares. For instance, $e_2$ has $\aff(e_2) = \{a_2, a_3\}$, $q(e_2) = 2$, and it approves of $a_1$ and $a_3$ for itself, $e_2$ for affiliate $a_2$, and $e_1$ and $e_3$ for affiliate $a_3$. On the right, we have a potential matching $\mu$ with an alternate matching $\mu'$. Note that $\mu'$ is the swapped matching of $\mu$ with respect to $\mathcal{T} = (a_3, a_2, a_2, e_2, e_1, e_1)$.}
    \label{fig:example}
    
\end{figure*}

%% file: survey.tex
\section{Evidence from a Human Experiment}\label{sec:human-experiments}

This survey strives to evaluate the applicability of our valuation function proposed for the \bamprob{}. More details of methods and results can be found in Appendix~\ref{appendix: protocol}.

In the survey, participants were asked to identify as a university in a \bamprob{} instance with an affiliated graduating student. Across multiple problem instances, the survey presented the user with binary preferences over relevant matches and five possible matchings. It then asked the users to rank the matchings. For each question, we found the preference profiles that emerge from our weighted valuation function when $\lambda \in\{\epsilon,1\}$ and then computed: (1) the chance of randomly selecting the profiles, and (2) empirical adherence to the profiles. These results are depicted in Table~\ref{tab:agreement}.

\begin{table}[]
  \centering
  \begin{adjustbox}{max width=.95\linewidth}
\begin{tabular}{@{}lcccccccc@{}}
\toprule
                                                                 \multicolumn{9}{c}{{\bf $\lambda=1$}}                                            \\ 
Scenario                                                                & 1      & 2     & 3     & 4      & 5     & 6     & 7     & 8      \\\midrule
{\it Random} & 10\% & 5\% & 5\% & 3\% & 100\% & 10\% & 10\% & 3\%\\
{\it Observed Unprimed}  & {\bf 41\%} & 31\% & 30\% &  {\bf 56\%} & 100\% & 42\% &  {\bf 25\%} & 30\%\\
{\it Observed Primed}  & {\bf 41\%} & {\bf 44\%} &  {\bf 35\%} & 55\% & 100\% &  {\bf 51\%} & 22\% &  {\bf 45\%}\\

 \midrule\midrule
 
                                                                  \multicolumn{9}{c}{{\bf $\lambda=\epsilon$}}                                            \\ 
Scenario                                                                & 1      & 2     & 3     & 4      & 5     & 6     & 7     & 8      \\\midrule
{\it Random} & 3\% & 1\% & 1\% & 3\% & 20\% & 3\% & 3\% & 1\%\\
{\it Observed Unprimed}  & 14\% & 15\% & 18\% & {\bf 56\%} & 30\% & 17\% & {\bf 16\%} & 17\%\\
{\it Observed Primed}  & {\bf 18\%} & {\bf 34\%} & {\bf 20\%} & 55\% & {\bf 37\%} & {\bf 28\%} & 10\% & {\bf 34}\%\\

 \midrule\midrule

\end{tabular}
\end{adjustbox}
  \caption{Percentage of respondents who followed a weighted valuation function with $\lambda=1$ and $\lambda=\epsilon$ for both primed and unprimed subjects. These are compared to the expected percentage if individuals were choosing randomly.}
  \label{tab:agreement}
\end{table}

Our results show that participants' ranking adherence to each valuation function is statistically significant, though not consistent. For a deeper quantitative analysis, see Appendix~\ref{appendix: protocol}. More qualitatively, participants expressed differing philosophies. Some participants were very direct with their strategies, even stating: ``{\it My needs first, then Ryan’s}'', where Ryan is the example affiliate. We see that our two valuation function versions align with this general strategy. On the other end of the spectrum, there were participants who were uncomfortable with the ability to express a preference over their affiliate's match. One participant said, ``{\it If Ryan doesn't get matched with my school, why would I care what others he matched with? Is it any of my business?}'' This indicates that there are clearly different strategies, but also different philosophical approaches to the affiliate matching problem.

%% file: theory.tex
\section{\bam{} Solved in Quadratic Time}\label{sec:theory}
We now introduce a quadratic (in the number of agents) algorithm, \textsf{SmartPriorityMatch}, to solve the \bamprob{} for general weights. We assume the inputs are provided as a set $A$ of applicants and $E$ of employers, where each agent reports all appropriate approval lists (i.e., lists of binary values). Let $n=|A|$ and $m=|E|$. All proofs are in the Appendix.

\begin{theorem}\label{thm:match}
\textsf{SmartPriorityMatch} solves the \bamprob{} in $O(nm)$ time for $\lambda \in [0,1]$.
\end{theorem}

Theorem~\ref{thm:match} is proved at the end of Appendix~\ref{appendix:sprm}. \textsf{SmartPriorityMatch}, shown in Algorithm~\ref{alg:spm} in Appendix~\ref{app:pseudocode}, effectively puts a ``priority level'' on each applicant-employer pair. For instance, a pair $(a,e)\in A\times E$ where $a\in\aff(e)$ and each agent is maximally interested in the match ($\pr_e^e(a)=1$, $\pr_e^a(e)=1$, and $\pr_a(e)=1$) is a ``highest'' priority edge. For each priority level, we construct bipartite graphs with partitions $A$ and $E$ where edges correspond to pairs of that priority level. The edge sets for the different priority levels (from highest to lowest priority) are as follows:

\begin{enumerate}
\item [$G_0$-] Edges between an $e\in E$ and $a\in\aff(e)$ if they have maximum interest for the match: $\pr_{e}^{e}(a) = 1$, $\pr_{e}^{a}(e) = 1$, and $\pr_{a}(e) = 1$.
\item [$G_1$-] Edges between an $e\in E$ and $a\in A\setminus\aff(e)$ if they are interested in each other: $\pr_{e}^{e}(a) = 1$ and $\pr_{a}(e) = 1$.

\item [$G_2$-] Edges between an $e\in E$ and $a\in\aff(e)$ if they are interested in each other for their own match: $\pr_e^e(a) = 1$ and $\pr_a(e) =1$.
\item [$G_3$-] Edges between an $e\in E$ and $a\in\aff(e)$ if $e$ is interested in itself for $a$ and $a$ is interested in $e$: $\pr_e^a(e) = 1$ and $\pr_a(e) =1$.
\end{enumerate}

Next, we would \textit{like} to run simple maximal $b$-matchings (i.e., many-to-many matchings) on these graphs in this order, decreasing the quotas as matches are made. Unfortunately, this method cannot ensure stability. Call this algorithm \textsf{PriorityMatch}. While this does not provide us with the desired results, it will set a strong foundation for \textsf{SmartPriorityMatch}.

\begin{lem}\label{lem:bad}
There exists a \bamprob{} instance where \textsf{PriorityMatch} may not find a stable matching.
\end{lem}

See Appendix~\ref{appendix:sprm} for a proof. Intuitively, \prm{}'s fault is that it is not sufficiently forward-looking. For instance, an employer $e$ that can match with one of two of its affiliates $a_1$ and $a_2$ in $G_0$ cannot greedily distinguish between the two. Therefore it could arbitrarily match with $a_1$, and $a_2$ could match with some other employer $e'$ in $G_1$. Perhaps $e$ likes the match $(e', a_1)$ and not $(e',a_2)$, in which case it should have matched with $a_2$ and let $a_1$ match with $e'$. This creates a blocking tuple $(e,e',e',a_2,a_1,a_1)$. This problem only arises when an employer might match with its affiliates who have the opportunity to match with other employers later on, which only happens on $G_0$. However, we find that \textsf{PriorityMatch} \textit{could} find a stable matching for these examples given a smart enough way to find the maximal $b$-matchings.

\begin{lem}\label{lem:prior}
\textsf{PriorityMatch} solves the \bamprob{} with parameter $\lambda\in[0,1]$ in $O(nm)$ time if it can ensure that for any potential blocking tuple $\mathcal{T} = (a,a',a'',e,e',e'')$ such that $a,a'\in\aff(e)$ and $a$ prefers the swapped matching of $\mu$ with respect to $\mathcal{T}$, then:
\[\pr_e^e(a') + \lambda\pr_e^a(e') + \lambda\pr_e^{a'}(e) \geq \pr_e^e(a) +  \lambda\pr_e^a(e) + \lambda\pr_e^{a'}(e'').\]
\end{lem}

See Appendix~\ref{appendix:sprm} for a proof. To achieve this, our maximal $b$-matchings must be dependent on lower-priority graphs. Instead of running the maximal $b$-matchings in order, we will use \textit{reserved matchings}, defined as follows.

\input{reserved_matching}

Our algorithm starts with a reserved matching on $G_1$, where reservations are used to  ensure we can still find a maximal matching on $G_0$ afterwards. Since each $a\in A$ may only be adjacent to $\aff^{-1}(a)$ in $G_0$, $G_0$ is a set of disjoint stars with centers $e\in E$. This lack of interference allows $e$ to match to any subset $S\subseteq N_{0}(e)$ of size exactly $|S| = \min(|N_{0}(e)|, q_{0}(e))$, where $N_{0}(e)$ is the neighborhood around $e$ in $G_0$ and $q_{0}(e)$ is the capacity of $e$ in $G_0$ (which is equivalent to its starting capacity). To ensure $e$ can do this after a reserved maximal $b$-matching in $G_1$, we must reduce the quota of $e$ in $G_1$ to $q_1 = q_{0}(e) - \min(|N_{0}(e)|, q_{0}(e))$ and ensure that at least $\min(|N_{0}(e)|, q_{0}(e))$ of its neighbors in $N_{0}(e)$ have at least one capacity remaining via a reservation on $N_{0}(e)$. Therefore, when we run the reserved maximal $b$-matching on $G_1$, we use affiliations $\mathcal{S}_1 = \{N_{0}(e),e\in E\}$ with reservations $r(N_{0}(e)) = \min(|N_{0}(e)|, q_{0}(e))$.

\input{graphs}

The algorithm thus works as follows: run a reserved maximal $b$-matching on $G_1$ and then proceed with the standard \prm{} process on $G_0$, $G_2$, and $G_3$. For more details, see the pseudocode of Algorithm~\ref{alg:spm} in Appendix~\ref{app:pseudocode}. It is not hard to see that the four resulting matchings are disjoint. We can show in our proofs that \textsf{SmartPriorityMatch} is in fact an intelligent implementation of \textsf{PriorityMatch}.

\begin{lem}\label{lem:smp}
\textsf{SmartPriorityMatch}'s output will always be equivalent to that of \textsf{PriorityMatch} with a specific maximal matching function.
\end{lem}

See Appendix~\ref{appendix:sprm} for a proof. Finally, we can show that \textsf{SmartPriorityMatch} satisfies the conditions posed in Lemma~\ref{lem:prior}. This concludes Theorem~\ref{thm:match}. We briefly note that this algorithm solves the problem for any weight $\lambda\in [0,1]$,\footnote{We additionally note that with a slight modification to edge priority, our algorithm could work for $\lambda\in[0,\infty)$.} however the algorithm itself does not depend on $\lambda$. Thus, there must exist a matching that is stable for all $\lambda$. We conjecture that increasing the value of $\lambda$ simply makes stability more difficult to achieve (i.e., for $1\geq\lambda>\lambda'\geq0$, a stable solution for $\lambda$ is also stable for $\lambda'$).

%% file: reserved_matching.tex
\begin{definition}\label{def:reservedmatch}
Consider a graph $G=(V,E,q,\mathcal{S},r)$, where $V,E,$ and $q$ are the standard $b$-matching parameters, $\mathcal{S}\subseteq 2^V$ is the set of \textbf{affiliations}, and $r:\mathcal{S}\to\mathbb{N}$ is a \textbf{reservation function} such that $r(S) \leq |S|$ for all $S\in\mathcal{S}$. A \textbf{reserved maximal $b$-matching} $\mu$ is a $b$-matching that is maximal under the additional constraint that for each $S\in\mathcal{S}$, there are at least $r(S)$ elements in $S$ that have not met their capacity: $|\{s\in S: |\mu(s)| < q(s)\}| \geq r(S)$.
\end{definition}

Consider an affiliation with 10 vertices, each with 100 capacity. The affiliation might have a reservation of 9 (of a max possible 10). We could match each vertex in the affiliation 99 times and one vertex 100 times. Only one vertex has reached its capacity, thus satisfying the reservation. We defer to Appendix~\ref{appendix:reserved} for a simple greedy solution to this problem.

%% file: graphs.tex
\begin{figure*}[h!]
  \centering
  \begin{subfigure}{.19\textwidth}
  \centering
  \vspace{0.3cm}
  \includegraphics[width=3cm]{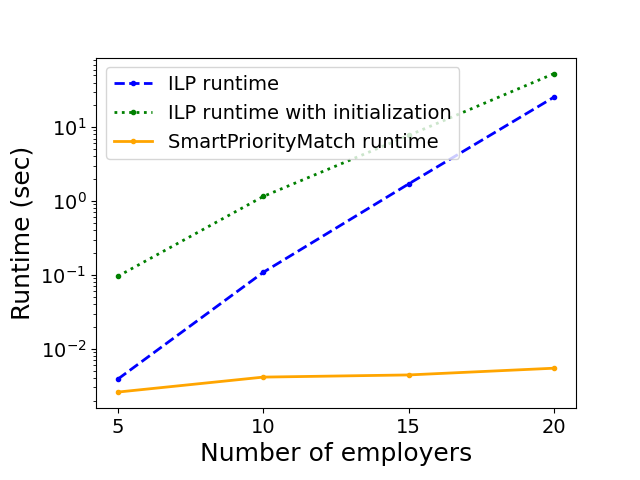}
  \captionsetup{width=2.3cm}
  \caption{Runtime vs ILP runtime for small values of $m$. We set $n/m=2$, $q=3$, and $t=0.5$.}
  \label{fig:sub1}
\end{subfigure}%
\begin{subfigure}{.19\textwidth}
  \centering
  \vspace{0cm}
  \includegraphics[width=3cm]{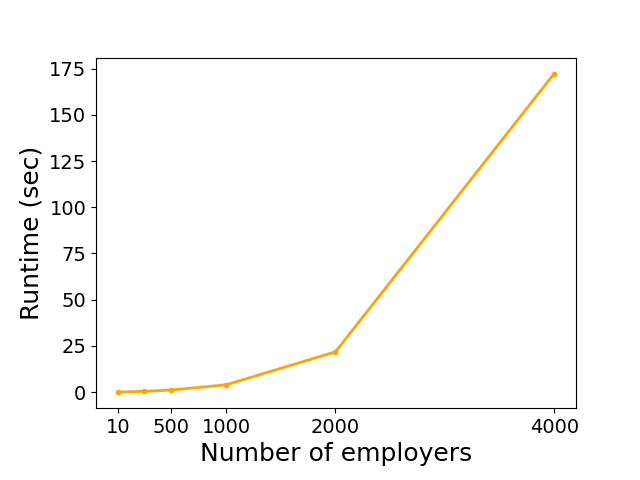}
    \captionsetup{width=2.3cm}
  \caption{Runtime for large values of $m$. We set $n/m=5$, $q=5$, and $t=0.5$.}
  \label{fig:sub2}
\end{subfigure}
\begin{subfigure}{.19\textwidth}
  \centering
  \vspace{0.3cm}
  \includegraphics[width=3cm]{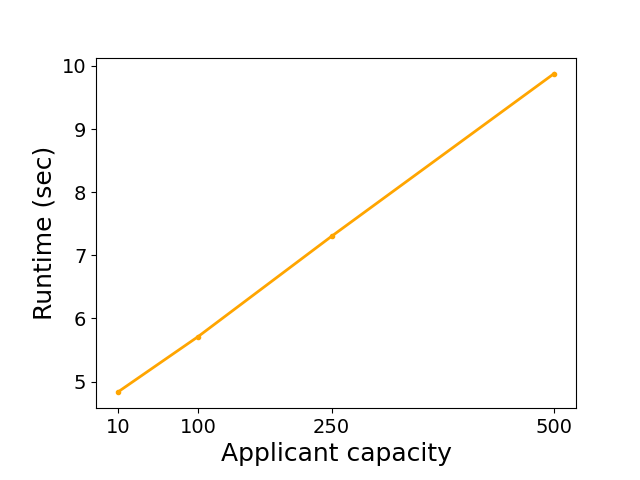}
    \captionsetup{width=2.3cm}
    \caption{Runtime over different capacities. We set $m=1000$, $n/m=5$, and $t=0.5$.}
  \label{fig:sub3}
\end{subfigure}
\begin{subfigure}{.19\textwidth}
  \centering
  \vspace{-0.05cm}
  \includegraphics[width=3cm]{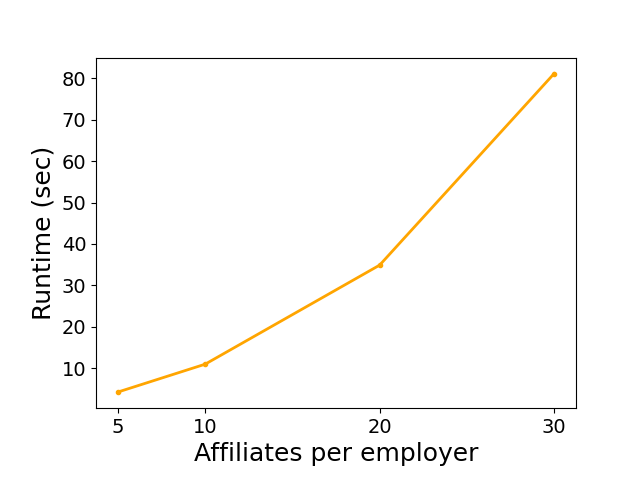}
    \captionsetup{width=2.3cm}
    \caption{Runtime over different values of $n/m$. We set $m=1000$, $q=5$, and $t=0.5$.}
  \label{fig:sub4}
\end{subfigure}
\begin{subfigure}{.19\textwidth}
  \centering
  \vspace{0.cm}
  \includegraphics[width=3cm]{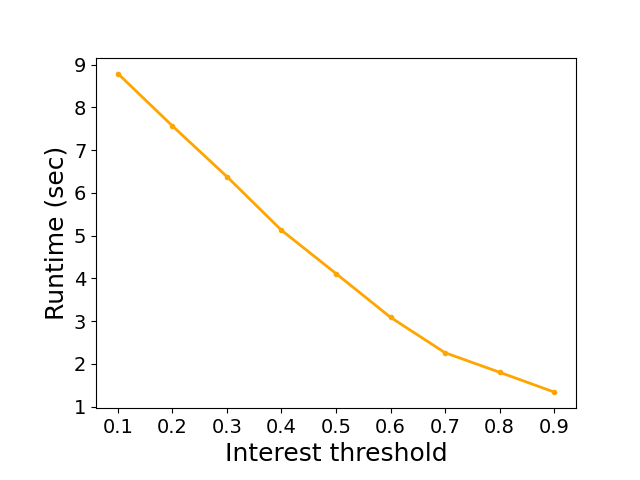}
    \captionsetup{width=2.3cm}
    \caption{Runtime over different approval thresholds. We set $m=1000$, $q=5$, and $n/m=5$.}
  \label{fig:sub5}
\end{subfigure}
   \caption{Runtime of \textsf{SmartPriorityMatch} while varying: number of employers ($m$), the capacities of the applicants ($q$), and the number of affiliates per employer ($n/m$). Note that the number of applicants is $m$ and the capacity of the employers is $q\cdot n/m$. }
  \label{fig:data}
\end{figure*}

%% file: experiments.tex
\section{Scalability Experiments}\label{sec:experiments} 

This section provides experimental validation for the polynomial-time scalability of \textsf{SmartPriorityMatch}, as analyzed in Theorem~\ref{thm:match}.  To the best of our knowledge, our model is new, so there is no direct benchmark from the literature.  Because of this, following the path of others (e.g., recently Cooper and Manlove~\shortcite{Cooper20:Algorithms}), we instead model our problem as an integer linear program (ILP) and compare against that baseline.  As in Cooper and Manlove~\shortcite{Cooper20:Algorithms} and other works, we also use that ILP as a ``safety check'' to ensure that our algorithmic approach and a general  mathematical-programming-based solution method align in their results. The formulation of the ILP and its proof can be found in Appendix~\ref{app:ilp}. It translates the set of potential blocking tuples (i.e., our blocking tuple search space) into ILP constraints. Since there are $O(n^3\cdot m^3)$ potential blocking tuples on $n$ applicants and $m$ employers, the ILP has $O(n^3\cdot m^3)$ constraints. %

To confirm the efficiency of \textsf{SmartPriorityMatch}, we compare it to the baseline ILP described in Appendix~\ref{app:ilp}. We use the same runtime experiments used by Tziavelis et al.~\shortcite{Tziavelis19:Equitable} adapted to the \bam{} setting.\footnote{That work addresses the traditional stable marriage problem and is thus not directly comparable to ours, but we adapt their experimental setup to our setting.}  We have four parameters: (1) $m$, the number of employers, (2) $n/m$, the number of affiliates per employer, (3) $q$, the capacity for each applicant, and (4) $t\in(0,1)$, a threshold parameter. This means that the number of applicants is $n$, and we let employer capacity be $q\cdot n/m$. We use Tziavelis et al.~\shortcite{Tziavelis19:Equitable}'s Uniform data, where we find a uniform random total ranking for each agent and we use the threshold parameter such that an agent with ranking $r$ is approved if $r > t\cdot n$. In other words, for each agent, we assign a preference of 1 to the top $100t$\% of its uniformly randomly ranked preferences. In Figure~\ref{fig:data}, we run 50 trials for each setting and take the average runtime.

We then vary $m$ from 5 to 20 and compare the performances of \textsf{SmartPriorityMatch} and the ILP (Figure~\ref{fig:sub1}) with $n/m=2$, $q=3$, and $t=0.5$ fixed. Since the ILP requires $O(n^3\cdot m^3)$ constraints, its runtime is very large for even small $n$. Due to our system's space constraints, we were only able to go up to $n=20$. We plotted the performance of the ILP with and without the time to initialize the ILP. We see that \textsf{SmartPriorityMatch} exhibits much better performance, particularly when we include the time to initialize the ILP itself.

Next we plot \textsf{SmartPriorityMatch}'s performance on larger sets, varying parameters one at a time. With $m$ from $10$ to $4000$ (Figure~\ref{fig:sub2}), we further support its scalability over the ILP. Varying $q$ from 5 to 500 (Figure~\ref{fig:sub3}), we see \textsf{SmartPriorityMatch} is dependent on capacity, but in practical ranges, it has less of an impact than varying $m$. With $n/m$ from 5 to 30 (Figure~\ref{fig:sub4}), we see that increasing $n/m$ has a significant impact on runtime. Finally, varying $t$ from 0.10 to 0.90, smaller thresholds appear to increase the runtime (i.e., when agents have a lower bar for expressing interest in other agents).

%% file: conclusions.tex
\section{Conclusions \& Future Research}\label{sec:conclusions}

We propose a new model, the \bamprob{}, that characterizes Dooley and Dickerson~\shortcite{Dooley20:Affiliate}'s affiliate matching problem under dichotomous preferences. Dichotomous, or approval-based, preferences are often more realistic for preference elicitation and their application to this model allows for stronger theoretical results. To rank matchings, we use a weighted function that computes agent matching valuations based off their and their affiliates' preferences. In a human survey, we support the real-world value use of the valuation function with different weights. We then develop (and prove) a quadratic time algorithm to solve the \bamprob{}, experimentally validating its efficiency against a baseline ILP.

This work could be extended by considering more general valuation functions, particularly by giving employers more freedom over the relative value of their and their affiliates' matches.  We may draw intuition from recent ``same-class'' preference extensions to the stable marriage problem such as the work of Kamiyama~\shortcite{Kamiyama20:Stable} or from stable matching work with constraints~\cite{Kawase20:Approximately}.  %
Similarly, we should consider concerns of fairness (other than stability).  Fair stable matching has a long history~\cite{Feder95:Stable,McDermid14:Sex-equal}, with hardness results~\cite{Gupta19:Balanced} for various forms of matching (e.g., with incomplete preferences~\cite{Cooper20:Algorithms} or other fairness constraints such as median-ranked assignment~\cite{Sethuraman06:Many}, equitable matching~\cite{Tziavelis19:Equitable}, procedural fairness~\cite{Tziavelis20:Fair}, etc.), many of which could be applied to the \bam{} setting. 

%% file: appendix.tex
\appendix
\section*{Appendix}

In the Appendix, we provide additional problem motivation, proofs, pseudocode, and survey details that were omitted in the body of the paper.

\input{motivation}

\input{model_proofs}

\input{survey_protocol}

\input{proofs3}

\input{spm_proofs}

\input{spm_pseudocode}

\input{ilp_proofs}

%% file: motivation.tex
\section{Problem Motivation (\S\ref{sec:intro})}

In our introduction, we mention multiple motivating examples for our model. In this section, we discuss them and their application to the model in further details.

\paragraph{Academic Faculty Interview Market} Our main motivating example is the academic faculty \textit{interview} marketplace. Here, graduating students applying for faculty positions are affiliated with their alma mater. Both students and universities indicate which agents on the other side of the market they are interested in interviewing with along with an interview slot capacity. Additionally, for each affiliate-university pair, the university indicates which universities they would like their affiliate to interview with, either for their own prestige or the well-being of the student. Note that this does not place a restriction on a student's matches based on their university's interest, but rather models another factor that may influence a university's preference over a complete interview matching.

The interview market, as opposed to the hiring market which motivated the model proposed by Dooley and Dickerson~\shortcite{Dooley20:Affiliate}, is more appropriate in this setting for two main reasons. First, dichotomous preferences seem more appropriate for interview matching, as interviews are intended to gauge interest on both sides, and thus the preference profile need not be refined. In faculty hiring, on the other hand, a preference profile maybe be inherently more complex than binary approval/disapproval. Second, the interview market lends itself to many-to-many matches, as both students and universities may desire multiple interviews, whereas the hiring market only generalizes to many-to-one matches, as students are only hired by one university. This simply better expresses the power of our solution which addresses the general many-to-many setting.

\paragraph{Playdate Matching} Another application is playdate matching. Consider a group of parents $\mathcal{P}$, each with an associated child, denoted by the set $\mathcal{C}$. Obviously, the affiliations are defined by parent-child relations, i.e., $\aff(p)$ is the child of $p\in\mathcal{P}$. These also denote either side of the market. A match between a parent $p\in \mathcal{P}$ and a child $c\in\mathcal{C}$ indicates that the child $c$ will go to $p$'s house to have a playdate with child $\aff(p)$. This can be a many-to-many matching, where quotas are how many playdates a child would like to go on and how many children a parent would like to host. Children can dis/approve of parents according to their interest in having a playdate with that child and/or going to their house, and parents dis/approve of children based off their interest in hosting the child and their own child's interest in the playdate. Additionally, parents' preferences over their child matches may come from whether or not the parent can drive a child to another parents' house or other related reasons. Instability indicates that a parent and child would forgo playdates to form a new playdate. Thus this is a nice application for the \bamprob{}.

\paragraph{Study Abroad} In the study abroad matching problem, we have a two-sided market consisting of current students interested in studying abroad and universities. Students are affiliated with the schools they attend, and they express interest or non-interest in other schools to go to study abroad. Schools express approval or non-approval for students that attend their program, as well as approval or non-approval of what study abroad programs they prefer to offer their own students. This also can be nicely modeled in terms of the \bamprob{}, where a stable solution ensures a university would never alter their accepted students in favor of other willing students in order to improve their valuation of the entire matching.

\paragraph{Student Project Allocation} Our next motivating example is the student project allocation market, where we use a quite similar (yet not identical) process to that of Manlove et al.~\cite{manlove2022student}. In this problem, there is a set of students $\mathcal{S}$, a set of lecturers $\mathcal{L}$, and a set of projects $\mathcal{P}$. Projects are proposed by lecturers, which defines a natural affiliation. Students express a preference over projects and lecturers express a preference over students for their affiliated projects. Note that we use approval-based preferences, whereas Manlove et al. use a combination of approval-based and ranked preferences. We believe it is reasonable in this application to elicit entirely dichotomous preferences.

To model this in the \bamprob{}, let the sides of the market be $\mathcal{P}$ and $\mathcal{S}\cap \mathcal{L}$ respectively. Note that we put $\mathcal{P}$ on the first side of the market because  is the side that will be affiliated with agents on the other side of the market. Let $\aff(s) = \emptyset$ for all $s\in \mathcal{S}$ and $\aff(l) = P_l$ where $P_l$ is the set of proposed projects by $l\in\mathcal{L}$. This forms a disjoint cover over $\mathcal{P}$. As projects do not have preferences over students, we simply set project preferences to approve of all students. The quota of a project is the number of students that may work on the project. Similarly, since faculty are not assigned to projects (this is a slight deviation from Manlove et al., where we assume lecturers are automatically assigned to all proposed projects), they must have zero quota. However, they exhibit preferences over the matches of their proposed projects. Finally, students also exhibit preferences over their matches with projects, and may have varying quotas depending on how many projects they are allowed to match with.

In this example, however, we note that stability is not entirely relevant. As lecturers are the only individuals with affiliations, and they have no quota, affiliations will actually not impact the stability of a matching. However, it does impact the overall value of a matching. Therefore, it may be interesting to explore other concepts of fairness in this model in light of this application.

\paragraph{Dog Breeding} Dog breeding is another problem that can be modeled using the \bamprob{}. In the dog breeding market, dog breeders have male and female dogs they would like to breed. In this application, we make the light simplifying assumption that the breeder who owns the female dog receives the offspring and the breeder who owns the male dog sells their services. To that end, the two sides of our market are as follows: in the first side, we have the male dogs, and on the second side, we have breeders, which encapsulates all female dogs they own. Clearly, the male dogs from a breeder are affiliated with that breeder.

Since male dogs do not have preferences, we simply assume male dogs approve of all possible matches (though they have some realistic capacity for matches). Breeders express their interest in male dogs they would like to purchase the services of (i.e., interest in their own matches) based off of the perceived breeding potential. They also express their preferences over breeders they would like their male dogs to service (i.e., interest in their affiliates' matches) based off of offered money, distance, etc.

Like the last example, stability in this model of dog breeding is not entirely compelling as dogs do not have agency to cheat as we describe in this model. However, as before, other notions of fairness may be of interest with respect to this application.

%% file: model_proofs.tex
\section{Model Definition Proofs (\S\ref{sec:model})}\label{appendix:model}

Here we prove the three propositions presented in Section~\ref{sec:model}. We start with Proposition~\ref{prop:matching}, which shows that a swapped matching of a matching with respect to a potential blocking tuple is still a matching. The proof is short and direct.

\begin{proof}[Proof of Proposition~\ref{prop:matching}]
We know $\mu'$ is a valid matching if no edge is matched across twice and no capacities are exceeded. The only formed matches are: $(a,e)$ and possibly $(a',e'')$ and $(a'',e')$. We know $(a,e)$ is unique as we require $e\notin \mu(a)$. Additionally, $e''$ and $a''$ are, by definition, not in $\mu(a')$ and $\mu(e')$ respectively. Therefore, since $\mu$ could not have duplicated matches, neither could $\mu'$. Both $a$ and $e$ lose and gain a match, and both $a'$ and $b'$ lose a match and possibly gain one match. Thus, their match sizes could not have increased, so they must not exceed their capacities. Finally, $a''$ and $e''$ might gain a match. This only happens if they are in $N^A_\mu$ and $N^E_\mu$ respectively, meaning they did not meet their quotas in $\mu$. Thus they could not exceed their quotas either.
\end{proof}

Next, in Proposition~\ref{def:block}, we show that we only have to consider potential blocking tuples in order to determine if a matching is stable. Furthermore, we can equate stability with the non-existence of a potential blocking tuple with specific preference profiles.

For intuition about why we can ignore some blocking tuples, we briefly show that there is a limit on the effect cheaters can have on the rest of the matching. For instance, assume $a\in A$ and $e\in E$ would like to cheat. In this instance, they can only break off matches to one $e'\in \mu(a)$ and $a'\in \mu(e)$ respectively. Thus, as a result of the cheating, only $e'$ and $a'$ could have new unused capacity. Then $e'$ and $a'$ may decide to match with each other, or they may decide to match with other individuals $a''\in A\setminus\mu(e')$ and $e''\in E\setminus\mu(a')$ with unmatched capacity respectively. For any other agents involved in the blocking tuple $\T$, their ability to match with each other is not a \textit{result} of $a$ and $e$ cheating.

Consider, for instance, some $a^*\in A\cap \T\setminus\{a,a',a''\}$ and $e^*\in E\cap \T\setminus\{e,e',e''\}$ that are in the tuple but not the aforementioned six affected agents. If $(a^*,e^*)$ is formed during the second step of the process from Definition~\ref{def:blocktup}, then that's simply because the two had additional capacity and preferred to match with each other. This is a valid notion of instability, however, it can be more simply captured by the tuple $(a^*,\gamma,\gamma,e^*,\gamma,\gamma)$ (recall that $\gamma$ is the empty agent with $\mathcal{E}=\{\gamma\}$), where all that happens is that $a^*$ and $e^*$ form a match.

\begin{proof}[Proof of Proposition~\ref{def:block}]
Consider a matching $\mu$ for an instance of the \bamprob{} and let $\T = (a_1,\ldots,a_k)$ be the blocking tuple for $\mu$. We show that if $\T$ is not a potential blocking tuple, then it implies that there is some tuple $\T_1$ with size $|\T_1| < \T$ is also a blocking tuple for $\mu$. This would then prove the first part of our results, that $\mu$ is unstable if and only if there exists a blocking tuple that is a potential blocking tuple.

Since $\T$ is not a potential blocking tuple, there is at least one agent $a^*\in \T$ who is not a cheater, is not cheated on, and does not match with an agent that is a cheater or cheated on. To see why, we consider multiple cases. First, if there are no cheaters, obviously no agents are cheaters or are cheated on, and therefore some agent in $\T$ must satisfy this.

Second, if there is one cheater, say without loss of generality the cheater is $a\in A$ who cheats on $e'\in E$, assume that all matches have at least one agent who is a cheater or who is cheated on. Since there is only one cheater ($a$) and one who is cheated on ($e'$), there can only be at most two such matches: $(a,e)$ for some $e\in E$ and $(e',a'')$ for some $a''\in A$. If $(a,e)$ is a match, then since it must be new to be involved in the blocking tuple, then $e\in E\setminus\mu(a)$. Similarly, if $(e',a'')$ is a match, $a''\in A\setminus\mu(E)$. Additionally, since $a''$ was not involved in cheating, it can only match with $e'$ if it had unmatched quota. Therefore $a''\in N_\mu^A\setminus\mu(e')$. There can be no other agents in $\T$ who form matches. Thus either there is some agent $a^*\in \T$ who does not match, and thus $a^*$ can be removed and we still have a smaller blocking tuple $\T_1$ (i.e., $a^*$ does not affect the final matching $\mu''$ since it does nothing), or $\T = (a,a'',e,e')$. In the latter case, $\T$ satisfies the conditions for a potential blocking tuple with $a',e'' \in \E$, which is a contradiction. Note that the argument is very similar if the cheater is $e\in E$ who cheats on some $a'\in A$.

Finally, we consider when there are two cheaters. By a similar argument as before, the only matches that involve a cheater or one who is cheated on are the match $(a,e)$ (which is required to be made in this case) for cheaters $a\in A$ and $e\in E\setminus \mu(a)$, the match $(a',e'')$ where $a'\in \mu(e)$ was cheated on and $e''\in E\setminus\mu(a')$, and the match $(a'',e')$ where $e'\in \mu(a)$ was cheated on and $a'' \in A\setminus\mu(e')$. Additionally, for $e''$ and $a''$ to be able to match with $a'$ and $e'$ respectively, they must have had unmatched quota after cheating occured. Therefore, they were either cheated on or had unmatched quota to start, meaning $e''\in (N_\mu^E\cup \{e'\})\setminus\mu(e')$ and $a''\in (N_\mu^A\cup \{e'\}\setminus\mu(a')$. Since $\T$ is not a potential blocking tuple, there must be some other agent $a^*\in \T\setminus\{a,a',a'',e,e',e''\}$. Otherwise, we use the same argument before to show we can create a smaller tuple $\T_1$ that is a blocking tuple for $\mu$. This concludes the first part of the proof.

At this point, we can assume $\T$ is a potential blocking tuple that is also a blocking tuple. Consider $\mu'$, the swapped matching of $\mu$ with respect to $\T$. Note that in the processed described in Definition~\ref{def:blocktup}, $\mu'$ is equivalent to the final matching. It is not hard to check that each of the six preference conditions in Proposition~\ref{def:block} correspond to the final conditions for the blocking tuple from Definition~\ref{def:blocktup}. For instance, $\mu\prec_a\mu'$ and $\mu\prec_e\mu'$ simply means $a$ and $e$ prefer the final matching to the starting matching. When $a'$ is an agent (i.e., $a\notin \E$), $\mu \prec_{a'} \mu\cup \{(a',e'')\}$ means $a'$ prefers to match with $e''$ over not in the context of the final matching. This is necessarily true by Definition~\ref{def:blocktup}, and the same argument holds for $a''$, $e'$, and $e''$. This concludes the forward direction of the proof. The reverse direction of the proof follows by simply observing that a blocking tuple with these preferences is necessarily a blocking tuple, so if one exists, then $\mu$ is clearly unstable.
\end{proof}

Finally, in the reserved maximal $b$-matching problem, it is fairly straightforward to show that the proposed \textsf{Greedy} algorithm achieves a maximal matching in $O(|E|)$ time on edge set $E$.

\begin{proof}[Proof of Proposition~\ref{prop:greedy-maximal-b-matching}]
Let $G= (V,E,q,\mathcal{S},r)$ be an instance of the reserved maximal $b$-matching problem and let $\mu$ be the matching returned by \textsf{Greedy}. At the start of \textsf{Greedy}, we clearly have a reserved $b$-matching. For every edge $e$, \textsf{Greedy} only adds $e$ if it does not break that the matching is a reserved $b$-matching. Thus it must always be a reserved $b$-matching throughout the algorithm.

Now we show $\mu$ is maximal. Consider some edge $e=(u,v)\notin\mu$, and let $\mu_e$ be the matching at the time $e$ was considered. Since $e$ was not added, its addition to $\mu_e$ would make it no longer a reserved $b$-matching. Therefore, it must break a quota or reservation constraint. Say it breaks a quota constraint for $v$ (wihtout loss of generality). Then $|\mu_e(v)| = q(v)$ since $\mu_e$ is a reserved $b$-matching but adding $e$ would break $v$'s quota. Since edges are only added throughout \textsf{Greedy}, $|\mu(v)| \geq |\mu_e(v)| = q(v)$, and since $\mu$ is a reserved $b$-matching, it must be that $|\mu(v)| = q(v)$. Therefore, adding $e$ to $\mu$ would make it no longer a reserved $b$-matching.

Otherwise, adding $e$ to $\mu_e$ would have broken a reservation constraint for some $S\in \mathcal{S}$. Adding $e$ to $\mu_e$, then, must fill the quota of at least one of $S$'s vertices. It clearly then must do this for one of its endpoints. If this happens to only one endpoint $v\in S$ (without loss of generality), then adding $e$ to $\mu_e$ makes $v$ meet its quota. Therefore $|\mu_e(v)| = q(v)-1$. As before, $|\mu(v)| \geq |\mu_e(v)| = q(v)-1$, so $|\mu(v)|\in\{q(v), q(v)-1\}$.  Additionally, since $S$'s reservation was broken by adding $e$ which only affected $v$'s quota in $S$, then $|S|-r(S)$ vertices in $S'\subseteq S\setminus\{v\}$ (with $|S'| = |S|-r(S)$) must have their quotas met in $\mu_e$. Since edges aren't removed, this is true in $\mu$ as well. Since $v\notin S'$, $v$ can't have met its quota in $\mu$, else $\mu$ would violate $S$'s reservation. Thus $|\mu(v)| = q(v)-1$. Adding $e$ to $\mu$, then, would make $v$ meet its quota, thus breaking the reservation for $S$. Thus, $e$ could not be added to $\mu$. In the final case, we consider if both $u$ and $v$ meet their quotas by adding $e$ to $\mu_e$. The analysis is essentially the same. This concludes the proof.
\end{proof}

%% file: survey_protocol.tex
\section{Survey Methods and Results (\S\ref{sec:human-experiments})} \label{appendix: protocol}
In this section, we provide real-world motivation for our valuation function in the \bamprob{}. To do this, we conducted an online survey that presented the \bamprob{} problems to participants. We find that the \bamprob{} induces behavior that shows an employer (in our survey, a university) may be willing to trade the quality of their match for that of their affiliate. We also find motivation for two polar versions of our valuation function: when $\lambda=1$ and $\lambda=\epsilon$. %

Our survey protocol can be found in the Appendix~\ref{appendix: protocol} and follows the work of Dooley and Dickerson~\shortcite{Dooley20:Affiliate}. 
We developed the survey protocol to answer our main research question: Do real-world participants follow the weighted valuation function and for which weights?

We hypothesize that individuals do follow our function, as we believe it is one of the most rational models for human behavior in the \bam{} setting. To explore this, we compute how often participants admit our tested weighted valuation function, considering both when $\lambda=1$ and $\lambda=\epsilon$, when presented with a specific \bam{} setting. 

We conducted the survey through a crowdsourcing platform, Cint, which connected us with English-speaking participants located in the United States. Our institution's IRB reviewed our survey structure and data-collection methods and determined it exempt and did not necessitate an IRB approval. After screening for setting comprehension, 203 participants completed the survey. Ten of those completed the survey too quickly (less than five minutes) and we excluded those responses. Each of the remaining 193 responses are included in our analysis below and we paid Cint \$3.05 for their time. The median response time of these 193 responses was 23 minutes. 

\subsection{Survey Design}

The participants were first introduced to the standard matching problem with three agents on each side of the market. They were primed to identify themselves as one of the universities. There was a matching-related test designed to filter out participants who were not paying attention to the content of the survey. The test consisted of three text-based questions in which the participant matched the text to a visual depiction of the corresponding matching scenario based off a dichotomous preference profile. Next, the concept of affiliates was introduced in the same matching setting with the five possible matches that involved their university and their affiliate. We then randomly primed the participant to believe that it was in their best interest to prioritize the match of their affiliate\footnote{Random priming, while not a main focus of our experiment, tests how differences in beliefs about one's own gain from the match of an affiliate would lead to different matching strategies}. Finally, we presented the participant with eight dichotomous scenarios and asked them to express their full preference over the possible matchings. These eight scenarios are fully detailed in the protocol in Appendix \ref{appendix: protocol}. 

\subsection{Results}

The main survey results are depicted in Table~\ref{tab:agreement}. To test if participants agreed with the valuation function, we first calculated the probability of uniform random responses resulting in a preference profile that follows the function for both parameters. We then compared this to the observed probability and found that with $p=0.05$ on a right-tailed alternative hypothesis of a binomial test that the increase in observed adherence to the valuation function for both parameterizations {\it is significant}.

     Additionally, of all participants, the median number of scenarios where they completely adhered to the $\lambda=1$ valuation function is 2. The same is true for the $\lambda=\epsilon$ valuation function. When $\lambda=1$, the median values between the primed and unprimed groups individually are 3 and 2 respectively.
When $\lambda=\epsilon$, the medians are 2 and 1 respectively.

Furthermore, 157 of the participants used the valuation function with $\lambda=\epsilon$ at least once and 164 used $\lambda=1$ at least once (excluding scenario 5). 
These results indicate that participants generally adhered to both valuation functions, but did not do so consistently throughout the survey. However, since their adherence was significantly higher than random, this suggests that there is structure in how the participants chose to adhere to the valuation functions. We pose for future research to design a survey instrument which investigates why a participant would choose whether or not to adhere to a particular valuation function.

We also observe that there is light evidence that our priming method was effective in inducing participants to follow the valuation function. Since this was not one of our central research questions with the survey ({\it does our priming method induce more deference to the affiliates' match}) we only mention that with further work, we could explore exactly how to prime the participants better. The purpose in performing the priming was to simulate the behavior of an admission faculty member. While this population is challenging to survey, our survey instrument does suggest that when considering the self-interest in your affiliate candidate's match, an agent may be more likely to follow either of the valuation function parameterizations.

\subsection{Complete Protocol}
This section includes the entire protocol used for the survey in Section \ref{sec:human-experiments}

{\textbf{Faculty Hiring Program}}

~

The design of this survey is aimed at understanding how you make decisions with different competing priorities. You will be exploring this concept in the setting of a hiring market for new faculty professors. The Survey will have two parts: (1) familiarization with faculty hiring, and (2) answering questions about your preferences. We begin with the familiarization part now.

Consider a hiring market such as this one with three applicants (Ryan,
Alex, and Taylor), and three universities (Bear Mountain, Littlewood,
and West Shores).

\includegraphics[width=.4\linewidth]{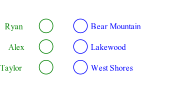}
\\
Imagine that \textbf{you are Bear Mountain University}, and you performed an evaluation of the applicants. You decided that you liked Alex more than you liked Taylor, and you liked Taylor more than you liked Ryan. 

\adjustbox{max width=\linewidth}{%
\begin{tabular}{c c c}
\toprule
{{Top Tier Candidates}} & {{Middle Tier Candidates}} & {{Bottom Tier
Candidates}}\\
Alex & Taylor & Ryan\\
\bottomrule
\end{tabular}
}

\hfill\break

You could depict that preference as:

\adjustbox{max width=\linewidth}{%
\begin{tabular}{c c c}
\toprule
{\bf{Alex}} & {\bf{Taylor}} & {\bf{Ryan}}\\
Interested & Not Interested & Interested\\
\bottomrule
\end{tabular}
}

Your preference over your possible student matches could then look like:

\textbf{{First Choice}}

\includegraphics[width=.3\linewidth]{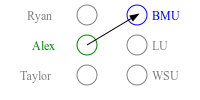}

\textbf{{Second Choice}}

\includegraphics[width=.3\linewidth]{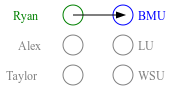}

\textbf{{Third Choice}}

\includegraphics[width=.3\linewidth]{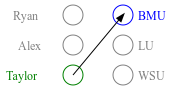}

But since you like Ryan too, you could also have the preferences:

\textbf{{First Choice}}

\includegraphics[width=.3\linewidth]{imgs/ryan.png}

\textbf{{Second Choice}}

\includegraphics[width=.3\linewidth]{imgs/alex.png}

\textbf{{Third Choice}}

\includegraphics[width=.3\linewidth]{imgs/taylor.png}

\noindent\makebox[\linewidth]{\rule{\linewidth}{0.4pt}}

To test your comprehension of the previous setting, can you now do the
matchings yourself? These are intuitive and should be easy to complete.~

~

Assume \textbf{you are Bear Mountain University}. Assume this time that after you review the applicants, you are interested in being matched with the candidates as follows:

\adjustbox{max width=\linewidth}{%
\begin{tabular}{c c c}
\toprule
{\bf{Alex}} & {\bf{Taylor}} & {\bf{Ryan}}\\
Interested & Not Interested & Interested\\
\bottomrule
\end{tabular}
}

Then what is your ranking of the following options?~

\textbf{Assuming you believe the above}, rank these outcomes from your
most preferred (1) to your least preferred (3).\\
\includegraphics[width=.3\linewidth]{imgs/alex.png}
\includegraphics[width=.3\linewidth]{imgs/ryan.png}
\includegraphics[width=.3\linewidth]{imgs/taylor.png}

~

Assume \textbf{you are Bear Mountain University}. Assume this time that
after you review the applicants, you place them in these tiers:

\adjustbox{max width=\linewidth}{%
\begin{tabular}{c c c}
\toprule
{\bf{Alex}} & {\bf{Taylor}} & {\bf{Ryan}}\\
Not Interested & Not Interested & Interested\\
\bottomrule
\end{tabular}
}

Then what is your ranking of the following options?~

\textbf{Assuming you believe the above}, rank these outcomes from your
most preferred (1) to your least preferred (3).\\
\includegraphics[width=.3\linewidth]{imgs/alex.png}
\includegraphics[width=.3\linewidth]{imgs/ryan.png}
\includegraphics[width=.3\linewidth]{imgs/taylor.png}

~

Assume \textbf{you are Bear Mountain University}. Assume this time that
after you review the applicants, you place them in these tiers:

\adjustbox{max width=\linewidth}{%
\begin{tabular}{c c c}
\toprule
{\bf{Alex}} & {\bf{Taylor}} & {\bf{Ryan}}\\
Not Interested & Interested & Not Interested\\
\bottomrule
\end{tabular}
}

Then what is your ranking of the following options?~

\textbf{Assuming you believe the above}, rank these outcomes from your
most preferred (1) to your least preferred (3).\\
\includegraphics[width=.3\linewidth]{imgs/alex.png}
\includegraphics[width=.3\linewidth]{imgs/ryan.png}
\includegraphics[width=.3\linewidth]{imgs/taylor.png}

\noindent\makebox[\linewidth]{\rule{\linewidth}{0.4pt}}

Awesome! Now onto the second part of the Survey. Since you understand
the basic faculty hiring setting, let us introduce another layer of
complexity.~

In faculty hiring, the applicants are affiliated with a university based
off where they earned their PhD. What this means is that universities
also care about where their student gets a job.

\textbf{Assume that you are Bear Mountain University and your student is
Ryan}. You then have the following five options of matchings:

You are matched with Ryan.

\includegraphics[width=.3\linewidth]{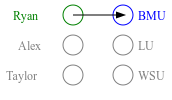}

You are matched with Alex; Ryan is matched with Littlewood University.

\includegraphics[width=.3\linewidth]{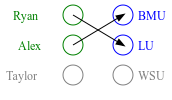}

You are matched with Taylor; Ryan is matched with Littlewood University.

\includegraphics[width=.3\linewidth]{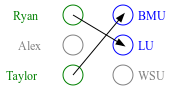}

You are matched with Alex; Ryan is matched with West Shores University.

\includegraphics[width=.3\linewidth]{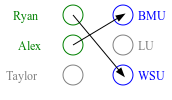}

You are matched with Taylor; Ryan is matched with West Shores University.

\includegraphics[width=.3\linewidth]{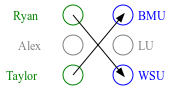}

In the remainder of the survey, we will ask you to express your preferences over these five options under different settings of which applicants you like best and what you think of the different universities.

\noindent\makebox[\linewidth]{\rule{\linewidth}{0.4pt}}

{\it Randomly assigned either of the two following prompts:}
\\

For the remainder of the survey, you will get to decide how you would like to balance your own interest in being matched with the best candidates possible, and where Ryan should be matched. There is no right answer -- it is up to you about how you balance these two things.

{\it or}

In faculty hiring, it is common to want your affiliate to be placed at a good university. This is often because you as a university will be perceived as a better university if your students get jobs at top tier schools.

So, keep in mind if Ryan is placed at a university you are interested in, then your university will be viewed better and could hire better candidates in the future. While you want Ryan to be matched with a good school, you must balance this priority with your competing priority that you want a good candidate. There is no right answer – it is up to you about how you balance these two things.
\\
We will ask you to express your preferences to 8 scenarios. When you are ready, please continue to Part 2.

\noindent\makebox[\linewidth]{\rule{\linewidth}{0.4pt}}

{\it Display this question 8 times with the interests as expressed in the enumerated list in Section \ref{sec:human-experiments}.}
\\
Assume that you are Bear Mountain University and Ryan is your student.~

Assume that you have evaluated the candidates and you decide that you are interested in hiring Alex, but you are not interested in hiring Ryan and Taylor.

\adjustbox{max width=\linewidth}{%
\begin{tabular}{c c c}
\toprule
{\bf{Alex}} & {\bf{Taylor}} & {\bf{Ryan (your student)}}\\
Interested & Not Interested &  Not Interested\\
\bottomrule
\end{tabular}
}

\hfill\break
Assume that you have the following beliefs about the schools, based off of the school’s ranking. You are interested in matching Ryan with Littlewood (LU) and your school (BMU), but you are not interested in Ryan being matched with West Shores (WSU).

\adjustbox{max width=\linewidth}{%
\begin{tabular}{c c c}
\toprule
{\bf{LU}} & {\bf{WSU}} & {\bf{BMU (your university)}}\\
Interested & Not Interested &  Interested\\
\bottomrule
\end{tabular}
}
\\

{\bf Assuming you believe the above}, rank these outcomes from your most preferred (1) to your least preferred (5).

\noindent\includegraphics[width=.3\linewidth]{imgs/qual_1.png}
\includegraphics[width=.3\linewidth]{imgs/qual_2.png}
\includegraphics[width=.3\linewidth]{imgs/qual_3.png}\\
\includegraphics[width=.3\linewidth]{imgs/qual_4.png}
\includegraphics[width=.3\linewidth]{imgs/qual_5.png}

Can you describe the procedure you used to rank these? Why did you use this procedure?

%% file: proofs3.tex
\section{Main Algorithm Proofs and Pseudocode (\S\ref{sec:theory})}\label{appendix:theory}

In this section, we present the proofs for our theoretical work. This includes proofs regarding the reserved maximal $b$-matching problem, \prm{} and \sprm{}. At the end, we present the pseudocode for the algorithm.

\subsection{The Reserved Maximal $b$-Matching Problem}\label{appendix:reserved}
We start by proposing a simple and efficien algorithm for solving the reserved maximal $b$-matching problem. This shows that \sprm{} can be implemented efficiently.

Let \textsf{Greedy} be a greedy algorithm for the reserved maximal $b$-matching problem, where we consider edges in an arbitrary order and greedily add them to the matching as long as they don't break any reservation or capacity constraints.

\begin{prop}\label{prop:greedy-maximal-b-matching}
\textsf{Greedy} solves reserved maximal $b$-matching in $O(\left|E\right|)$ time for $E$ edges.
\end{prop}

\begin{proof}
Let $G= (V,E,q,\mathcal{S},r)$ be an instance of the reserved maximal $b$-matching problem and let $\mu$ be the matching returned by \textsf{Greedy}. At the start of \textsf{Greedy}, we clearly have a reserved $b$-matching. For every edge $e$, \textsf{Greedy} only adds $e$ if it does not break that the matching is a reserved $b$-matching. Thus it must always be a reserved $b$-matching throughout the algorithm.

Now we show $\mu$ is maximal. Consider some edge $e=(u,v)\notin\mu$, and let $\mu_e$ be the matching at the time $e$ was considered. Since $e$ was not added, its addition to $\mu_e$ would make it no longer a reserved $b$-matching. Therefore, it must break a quota or reservation constraint. Say it breaks a quota constraint for $v$ (wihtout loss of generality). Then $|\mu_e(v)| = q(v)$ since $\mu_e$ is a reserved $b$-matching but adding $e$ would break $v$'s quota. Since edges are only added throughout \textsf{Greedy}, $|\mu(v)| \geq |\mu_e(v)| = q(v)$, and since $\mu$ is a reserved $b$-matching, it must be that $|\mu(v)| = q(v)$. Therefore, adding $e$ to $\mu$ would make it no longer a reserved $b$-matching.

Otherwise, adding $e$ to $\mu_e$ would have broken a reservation constraint for some $S\in \mathcal{S}$. Adding $e$ to $\mu_e$, then, must fill the quota of at least one of $S$'s vertices. It clearly then must do this for one of its endpoints. If this happens to only one endpoint $v\in S$ (without loss of generality), then adding $e$ to $\mu_e$ makes $v$ meet its quota. Therefore $|\mu_e(v)| = q(v)-1$. As before, $|\mu(v)| \geq |\mu_e(v)| = q(v)-1$, so $|\mu(v)|\in\{q(v), q(v)-1\}$.  Additionally, since $S$'s reservation was broken by adding $e$ which only affected $v$'s quota in $S$, then $|S|-r(S)$ vertices in $S'\subseteq S\setminus\{v\}$ (with $|S'| = |S|-r(S)$) must have their quotas met in $\mu_e$. Since edges aren't removed, this is true in $\mu$ as well. Since $v\notin S'$, $v$ can't have met its quota in $\mu$, else $\mu$ would violate $S$'s reservation. Thus $|\mu(v)| = q(v)-1$. Adding $e$ to $\mu$, then, would make $v$ meet its quota, thus breaking the reservation for $S$. Thus, $e$ could not be added to $\mu$. In the final case, we consider if both $u$ and $v$ meet their quotas by adding $e$ to $\mu_e$. The analysis is essentially the same. This concludes the proof.
\end{proof}

\subsection{\prm{} Proofs}\label{appendix:sprm}

In this section, we address all proofs regarding \prm{}. First, we introduce a new lemma that will simply show a useful property that we use throughout these proof. The high level idea is that if matching $\mu'$ is a swapped matching of $\mu$, agents only care about the part of their matches (and possibly their affiliate's matches) that change to decide which matching they like better. Specifically, we give conditions for an $a\in A$ and $e\in E$ where $a\notin\mu(e)$ but $a\in\mu'(e)$ would \textit{not} strongly prefer the new match to the old one. If this can be shown for all possible ways to do a pairwise swap to match $a$ and $e$, then they cannot form a blocking tuple. This is crucial to our proof.

\begin{lem}\label{lem:prefs}
Consider the \bamprob{}. Let $\mathcal{T} = (a,a',a'',e,e',e'')$ be a potential blocking tuple for a matching $\mu$, and $\mu'$ be the swapped matching of $\mu$ with respect to $\mathcal{T}$. Then both of the following hold:
\begin{enumerate}
\item If $\pr_a(e) \leq \pr_a(e')$, then $a$ cannot prefer $\mu'$ to $\mu$.
\item Let $I_a$, $I_{a'}$, and $I_{a''}$ be the respective indicators that $a,a',$ and $a''$ are in $\aff(e)$. Let $J_{a'',a'}$ be the indicator that $a''\neq a'$. If 
\begin{align*}
\pr_e^e(a') + \lambda I_a&\pr_e^a(e') + \lambda I_{a'} \pr_e^{a'}(e)
\\\geq\pr_e^e(a) &+ \lambda I_a\pr_e^{a}(e) + \lambda I_{a'}\pr_e^{a'}(e'') \\&+ \lambda I_{a''}J_{a'',a'}\pr_e^{a''}(e'),
\end{align*}
then $e$ cannot prefer $\mu'$ to $\mu$.
\end{enumerate}
\end{lem}

\begin{proof}
For 1:
\begin{align*}
\agg_a(\mu') =& \pr_a(e) + \sum_{e^*\in \mu'(a)\setminus e} \pr_a(e^*)
\\\leq& \pr_a(e') + \sum_{e^*\in \mu(a)\setminus e'} \pr_a(e^*)\tag{1}\label{line:prefs1}
\\=&\agg_a(\mu),
\end{align*}
where line~\ref{line:prefs1} comes from the fact that the two summations sum over the same matches of $a$, and $\pr_a(e) \leq \pr_a(e')$ (as given). Thus, $a$ does not strongly prefer $\mu'$ to $\mu$. This completes the first part of the Lemma. We will now show 2, for $b_1,\ldots,b_{|\aff(a)|}\in\aff(a)$:
\begin{align*}
\agg_e(\mu') =& \pr_e^e(a) + \sum_{a^*\in\mu'(e)\setminus\{a\}} \pr_e^e(a^*) 
\\&+ \lambda I_{a} \left(\pr_e^{a}(e) + \sum_{e^*\in\mu'(a)\setminus\{e\}} \pr_e^{a}(e^*)\right) 
\\&+ \lambda I_{a'} \left(\pr_e^{a'}(e'') + \sum_{e^*\in\mu'(a')\setminus\{e''\}} \pr_e^{a'}(e^*)\right) \\&+\lambda  I_{a''}J_{a'',a'} \left(\pr_e^{a''}(e') + \sum_{e^*\in\mu'(a'')\setminus\{e'\}} \pr_e^{a''}(e^*)\right) 
\\&+ \lambda \sum_{a^*\in \aff(e)\setminus\{a,a',a''\}} \sum_{e^*\in \mu'(a^*)} \pr_e^{a^*} (e^*)
\\\leq& \pr_e^e(a') + \sum_{a^*\in\mu(e)\setminus\{a'\}} \pr_e^e(a^*) 
\\&+ \lambda I_{a} \left(\pr_e^{a}(e') + \sum_{e^*\in\mu(a)\setminus\{e'\}} \pr_e^{a}(e^*)\right) 
\\&+ \lambda I_{a'} \left(\pr_e^{a'}(e) + \sum_{e^*\in\mu(a')\setminus\{e\}} \pr_e^{a'}(e^*)\right) \\&+\lambda  I_{a''}J_{a'',a'} \left(\sum_{e^*\in\mu(a'')} \pr_e^{a'}(e^*)\right) 
\\&+ \lambda \sum_{a^*\in \aff(e)\setminus\{a,a',a''\}} \sum_{e^*\in\mu(a^*)}\pr_e^{a^*} (e^*)\tag{2}\label{line:prefs2}
\\=& \agg_e(\mu)
\end{align*}
where line~\ref{line:prefs2} comes from the fact that all the summations are over the same matches, and otherwise the terms we pull out compose the provided inequality. When we pull out the summation for $a''$, note that this is only something we can pull out if $a''\neq a'$, because otherwise we already pulled it out as the $a'$ summation. This is why we multiply it by $J_{a'',a'}$.
\end{proof}

We now introduce another useful lemma that implies that in any blocking tuple with respect to the matching found by \prm, $a$ cannot be matched. Equivalently:

\begin{lem}\label{lem:tup}
Let $\mu$ be the resulting matching from \textsf{PriorityMatch}. Then any potential blocking tuple $\T = (a,a',a'',e,e',e'')$ that blocks $\mu$ must satisfy $e',a''\in\mathcal{E}$ and $a',e''\notin \mathcal{E}$.
\end{lem}

\begin{proof}
Fix a blocking tuple $\T = (a,a',a'',e,e',e'')$. Let $\mu'$ be the swapped matching of $\mu$ with respect to $\T$. Assume for contradiction that $e'\notin\mathcal{E}$. That means $e'\in\mu(a)$ by the definition of the blocking tuple. Notice that every edge $(a^*,e^*)$ in every subgraph in \prm{} satisfies $\pr_{a^*}(e^*)=1$. Since all matches are selected from these edge and $(a,e')$ was matched by \prm{}, this implies $\pr_a(e')=1$. By Lemma~\ref{lem:prefs}, $a$ cannot prefer $\mu'$ to $\mu$. This contradicts that $\T$ is a blocking tuple. This proves $e'\in\mathcal{E}$. By the definition of the blocking tuple, we then know $a''\in\mathcal{E}$.

Next, consider the case when $a'\in\mathcal{E}$ or $e''\in\mathcal{E}$. By the definition of the potential blocking tuple, one implies the other, thus $a',e''\in\mathcal{E}$. This means $a$ and $e$ both are not matched to capacity and they simply use this to match with each other. Since each must have an interest in each other to form a blocking tuple, (i.e., $\pr_{a}(e)=1$ and $\pr_e^e(a)=1$ or $\pr_e^a(e)=1$), that means $(a,e)$ is an edge in one of the \prm{} graphs. Since they never reached capacity, they must have then matched during the maximal matching in that graph. This contradicts that $e\notin\mu(a)$.
\end{proof}

Combining these two lemmas yields the following useful lemma:

\begin{lem}\label{lem:prefs2}
Consider the \bamprob{}. Let $\mathcal{T} = (a,a',a'',e,e',e'')$ be a potential blocking tuple for a matching $\mu$, and $\mu'$ be the swapped matching of $\mu$ with respect to $\mathcal{T}$. Let $I_a$ and $I_{a'}$ be the respective indicators that $a$ and $a'$ are in $\aff(e)$. If 
\begin{align*}
\pr_e^e(a')& + \lambda I_{a'} \pr_e^{a'}(e) \geq\pr_e^e(a) + \lambda I_a\pr_e^{a}(e) + \lambda I_{a'}\pr_e^{a'}(e''),
\end{align*}
then $e$ cannot prefer $\mu'$ to $\mu$.
\end{lem}

\begin{proof}
Let $I_{a''}$ be the indicator that $a''\in\aff(a)$ and $J_{a'',a'}$ be the indicator that $a''\neq a'$. By Lemma~\ref{lem:tup}, $I_{a''}=0$ and $\pr_e^a(e')=0$ since $e',a''\in\mathcal{E}$. The two following equations hold:
\begin{align*}
\pr_e^e&(a') + \lambda I_{a'} \pr_e^{a'}(e)\\=&\pr_e^e(a') + \lambda I_a\pr_e^a(e') + \lambda I_{a'} \pr_e^{a'}(e),
\\\pr_e^e&(a) + \lambda I_a\pr_e^{a}(e) + \lambda I_{a'}\pr_e^{a'}(e'')\\=&\pr_e^e(a) + \lambda I_a\pr_e^{a}(e) + \lambda I_{a'}\pr_e^{a'}(e'') + \lambda I_{a''}J_{a'',a'}\pr_e^{a''}(e').
\end{align*}
Substituting these into each side for the second inequality in Lemma~\ref{lem:prefs} gives the desired result.
\end{proof}

Note that this is a more useful version of the second inequality of Lemma~\ref{lem:prefs}. Since we will need a version of both inequalities of Lemma~\ref{lem:prefs}, we will now canonically refer to Lemma~\ref{lem:prefs} to refer to its first inequality, and Lemma~\ref{lem:prefs2} for the simpler version of the second inequality. We now prove the lemma defining when \prm{} works.

\begin{proof}[Proof of Lemma~\ref{lem:prior}]
Obviously, as the quotas always reflect the maximum remaining space any agent has for  matches, this creates a valid matching. We now show this is stable.

Consider $\mu$, the output of \textsf{PriorityMatch}. Assume that it satisfies the preconditions of this lemma: for any potential blocking tuple $\mathcal{T} = (a,a',a'',e,e',e'')$ such that $a,a'\in\aff(e)$ and $a$ prefers the swapped matching of $\mu$ with respect to $\mathcal{T}$, then:

\[\pr_e^e(a') + \lambda\pr_e^{a'}(e) \geq \pr_e^e(a) +  \lambda\pr_e^a(e) + \lambda\pr_e^{a'}(e''). \tag{1}\label{eq:precond}\]

By Lemma~\ref{lem:tup}, we can ignore the $e'$ and $a''$ parameters as $e',a''\in \mathcal{E}$, so they are not involved in the blocking tuple. Thus our tuple is effectively reduced to $\mathcal{T} = (a,a',e,e'')$. Assume for contradiction that $\mathcal{T}$ is a blocking tuple. Let $\mu'$ be the swapped matching of $\mu$ with respect to $\mathcal{T}$. Note that to use Inequality~\ref{eq:precond}, it is sufficient to show that $a,a'\in \aff(e)$ and $\mu' \succ_a \mu$ (i.e., the precondition of the lemma). The latter is satisfied because we assume $\T$ is a blocking tuple. Additionally, if we were to satisfy this, then $I_a=I_{a'}=1$, which means $\pr_e^e(a') + \lambda\pr_e^{a'}(e) = \pr_e^e(a') + \lambda I_{a'}\pr_e^{a'}(e)$ and $\pr_e^e(a) +  \lambda\pr_e^a(e) + \lambda\pr_e^{a'}(e'') = \pr_e^e(a) +  \lambda I_a\pr_e^a(e) + \lambda I_{a'}\pr_e^{a'}(e'')$. Plugging these into Inequality~\ref{eq:precond}, we get:

\begin{align*}
\pr_e^e(a')& + \lambda I_{a'} \pr_e^{a'}(e) \geq\pr_e^e(a) + \lambda I_a\pr_e^{a}(e) + \lambda I_{a'}\pr_e^{a'}(e''),
\end{align*}

This satisfies Lemma~\ref{lem:prefs2}, which implies $e$ cannot prefer $\mu'$ to $\mu$. Therefore, $a$ and $a'$ cannot both be in $\aff(e)$ simultaneously. 

Given that $e',a''\in\mathcal{E}$  and $a',e''\notin \mathcal{E}$ by Lemma~\ref{lem:tup}, it must be the case, then, that $a\in N_\mu^A$ is simply filling unmet capacity while $e$ has met its capacity and is therefore dropping its match with some $a'\in A$ in order to match with $a$. Then $a'$ may or may not match with another employer (depending on if $e''\in\mathcal{E}$ or $e''\in E$). We can then do a case by case analysis based off the fact that $a$ and $a'$ cannot both be in $\aff(e)$ (note that we say $\mu_i$ is the maximal matching found by \prm{} in $G_i$):

\begin{enumerate}
\item If $a\in\aff(e)$: Then $I_a=1$ and $a'\notin\aff(e)$ so $I_{a'} = 0$. Since $a'\notin\aff(e)$, $(a',e)$ can only have been in $G_1$. Since it was matched by \prm{}, it must have been in $G_1$. Therefore, $\pr_e^e(a')=1$. Thus, on the lefthand side of Lemma~\ref{lem:prefs2}:
\[\pr_e^e(a') + \lambda I_{a'}\pr_e^{a'}(e) = 1\]
On the righthand side: 
\[\pr_e^e(a) + \lambda I_a\pr_e^{a}(e) + \lambda I_{a'}\pr_e^{a'}(e'') = \pr_e^e(a) + \lambda \pr_e^a(e)\]
Since $\mathcal{T}$ is a blocking tuple by assumption, then by Lemma~\ref{lem:prefs2} we must have $\pr_e^e(a) + \lambda \pr_e^a(e) > 1$. This can only happen if $\pr_e^e(a) = \pr_e^a(e) = 1$. Since Lemma~\ref{lem:prefs} implies $\pr_a(e) = 1$, $(e,a)$ must have been an edge in $G_0$. Note also that since $a'\notin\aff(e)$, $(e,a')\notin G_0$. Thus $a$ did not meet its quota in $G_0$ (it never did) and $e$ must have had remaining quota after $\mu_0$. This contradicts the maximality of $\mu_0$.
\item Else if $a'\in\aff(e)$: Then $I_{a'} = 1$ and $a\notin \aff(e)$ so $I_a=0$. We start by simplifying the left and righthand side of  Lemma~\ref{lem:prefs2}:
\begin{align*}
\pr_e^e(a') + \lambda I_a\pr_e^{a'}(e) &= \pr_e^e(a') + \lambda \pr_e^{a'}(e) 
\\\pr_e^e(a) + \lambda I_a\pr_e^{a}(e) + \lambda I_{a'}\pr_e^{a'}(e'') &= \pr_e^e(a) + \lambda \pr_e^{a'}(e'')
\end{align*}
For $\mathcal{T}$ to be a blocking tuple, Lemma~\ref{lem:prefs2} implies:
\[\pr_e^e(a') + \lambda \pr_e^{a'}(e) < \pr_e^e(a) + \lambda \pr_e^{a'}(e'')\]
The lefthand side can be 0, $\lambda$, 1, or $1+\lambda$. It obviously cannot be 0, else $\pr_e^e(a') = \pr_e^{a'}(e) = 0$, so $(e,a')$ could not have been an edge in any graph, and therefore could not have been selected by \prm{}. It also cannot be $1+\lambda$, since this is the maximum value of the righthand side.

Consider if $\pr_e^e(a') + \lambda \pr_e^{a'}(e) \in \{1,\lambda\}$. Then $\pr_e^{a'}(e) = 0$ and $\pr_e^e(a') = 1$ or the reverse, so $(e,a')$ was an edge in $G_2$ or $G_3$ respectively, implying that $e$ had unmatched quota until at least either $\mu_2$ or $\mu_3$ occurred respectively (i.e., it had unmatched quota in $G_2$ or $G_3$). Since $a$ always had unmatched quota during \prm{}, it cannot have appeared in any graph prior to $G_2$ or $G_3$ respectively, else \prm{} would have matched $e$ and $a$. In the first case, which occurs when the lefthand side is $1$, we know it cannot be the case that $\pr_e^e(a) = \pr_e^{a'}(e) = 1$ (since that would put $(e,a')$ in $G_0$), thus the righthand side is bound above by 1 so the inequality cannot hold. In the second case, which occurs when the lefthand side is $\lambda$, we know $\pr_e^e(a) = 0$ (otherwise $(e,a')$ would be in $G_0$ or $G_2$), thus the righthand side is bound above by $\lambda$ so the inequality cannot hold. In either case, this is a contradiction.
\item Else: $a,a'\notin\aff(e)$, so $I_a=I_{a'}=0$. Again, we simplify the left and righthand side of Lemma~\ref{lem:prefs2}:
\begin{align*}
\pr_e^e(a') + \lambda I_a\pr_e^{a'}(e) &= \pr_e^e(a')
\\\pr_e^e(a) + \lambda I_a\pr_e^{a}(e) + \lambda I_{a'}\pr_e^{a'}(e'') &= \pr_e^e(a)
\end{align*}
Since \prm{} matched $(e,a')$, $(e,a')$ must have been in some graph. For this to happen when $a'\notin\aff(e)$, it must be the case that $\pr_e^e(a') = 1$. Thus the lefthand side becomes 1, which is also an upper bound on the righthand side. Thus Lemma~\ref{lem:prefs2} holds, contradicting that $\mathcal{T}$ is a potential blocking tuple.
\end{enumerate}
Thus we have contradicted the possibility of having a blocking tuple, so our algorithm produces a stable marriage.
\end{proof}

Finally, we prove the weakness of \textsf{PriorityMatch} in the prioritized setting.

\begin{proof}[Proof of Lemma~\ref{lem:bad}]
Consider a market with two agents on each side: $A=\{a_1,a_2\}$ and $E=\{e_1,e_2\}$. We let $\aff(e_1) = \{a_1,a_2\}$ and $\aff(e_2) = \emptyset$. All possible preferences are set to one \textit{except} $\pr_{e_1}^{a_2}(e_2) = 0$, $\pr_{e_2}^{e_2}(a_2)=0$, and $\pr_{a_2}(e_2) = 0$. All quotas are 1. This means that $G_0$ contains edges $(a_1,e_1)$ and $(a_2,e_1)$. Since $e_1$ has a quota of 1, it must select only one edge. In an arbitrary maximal matching, it might select $(a_1,e_1)$. All other graphs, at this point, will either be empty, or only have edges with at least one 0-quota endpoint. Thus $a_2$ and $e_2$ will not be matched (though, we could imagine just arbitrarily matching the rest at the end and our argument will hold). However, note that $a_2$ likes $e_1$ and dislikes $e_2$. Similarly, $e_1$ would prefer for its affiliate $a_1$ to be matched with $e_2$ than $a_2$ be matched with $e_2$. Clearly, $a_2$ and $e_1$ would prefer to be matched and allow $a_1$ and $e_2$ to be matched. 
\end{proof}

%% file: spm_proofs.tex
\subsection{\sprm{} Proofs}

Next, we must show that \textsf{SmartPriorityMatch} is effectively an implementation of \textsf{PriorityMatch} with a smarter algorithm for the matchings. In order to do this, we need to show that each matching could have been generated by \textsf{PriorityMatch} given the prior matches. We show this one step at a time. Note there are slight differences between the graphs for \textsf{PriorityMatch} and \textsf{SmartPriorityMatch} in the quotas of the graphs and reservations. To clarify, for instance, the quotas in the first graph for each algorithm, we use $q_0$ for \prm{} and $q_0'$ for \sprm{}.

\begin{lem}\label{lem:smp0}
Let $G_0 = (V_0, E_0, q_0)$ be the first subgraph considered by \textsf{PriorityMatch} with capacities $q_0$. Let $\mu_0$ be the matching \textsf{SmartPriorityMatch} finds on $G_0$. Then $\mu_0$ is a maximal matching on $G_0$.
\end{lem}

\begin{proof}
Let $G_1 = (V_1, E_1, q_1)$ be the second subgraph considered by \prm{}. Before \textsf{SmartPriorityMatch} finds $\mu_0$, it runs a reserved maximal matching $\mu_1$ on $(V_1, E_1, q_1', \mathcal{S}, r)$, where $q_1'$ are the capacities used by \sprm{} on $V_1$, $\mathcal{S}$ are the affiliations, and $r$ are the reservations. Next, it finds some maximal matching $\mu_0$ on $(V_0, E_0, q_0')$, where $q_0'$ are the capacities used by \sprm{} on $V_0$. Since we use the same edge set and $q_0(v)' = q_0(v) - |\mu_1(v)| \leq q_0(v)$ for each $v\in V$, $\mu_0$ is a valid matching on $G_0$.

Assume for contradiction there is some $(e,a)\in E_0\setminus \mu_0$ such that $\mu_0\cup\{(e,a)\}$ is a valid matching in $G_0$. Consider the topology of $(V_0, E_0)$. It is a set of disjoint stars connecting each employer (the center) to a subset of its affiliates. In $G_0$, we then know $e$ is going to match to some subset of its neighborhood. Since the stars are all disjoint and all star leaves have capacity at least one, then in a maximal matching, $e$ can and must match to any $\min(|N_{0}(e)|, q_0(e))$-sized subset of its neighbors during \prm{}.

Since $\mu_0\cup\{(e,a)\}$ is a valid matching in $G_0$, that means there is some set of matches $S$ that involve $e$ with $(e,a) \in S$ such that $\mu_0(e)\cup S$ is maximal (i.e., you cannot add any more matches to $e$ without breaking some agent's quota). By our argument from before, $|\mu_0(e)\cup S| = \min(|N_{0}(e)|, q_0(e))$, therefore $|\mu_0(e)| < \min(|N_{0}(e)|, q_0(e))$. Since the structure of graph $(V_0,E_0,q_0')$ is the same as $(V_0,E_0,q_0)$, $\mu_0(e)$ must abide by the same properties to be maximal, notably that $|\mu_0| = \min(|N_{0}'(e)|, q_0'(e))$ where $N_0'(e) = \{a\in N_0(e): q_0'(a) > 0\}$. Therefore, $\min(|N_{0}(e)|, q_0(e)) > \min(|N_{0}'(e)|, q_0'(e))$.

We start by considering $|N_0'(e)|$. It is equivalent to the number of agents $a\in N_0(e)$ who did not match to their quota in $\mu_1$. Since $\mu_1$ is a reserved maximal b-matching with  $N_0(e)\in \mathcal{S}$ with reservation $r(N_0(e)) = \min(|N_0(e)|, q_0(e))$, this number is at least $\min(|N_0(e)|, q_0(e))$. Therefore, $|N_0'(e)| \geq \min(|N_0(e)|, q_0(e))$.

Thus for $\min(|N_{0}(e)|, q_0(e)) > \min(|N_{0}'(e)|, q_0'(e))$ to hold, it must be the case that $q_0'(e) < \min(|N_0(e)|, q_0(e))$. Note that we set $q_0'(e) = q_0(e) - |\mu_1(e)|$. Thus $q_0(e) - |\mu_1(e)| < \min(|N_{0}(e)|, q_0(e))$, and so $q_0(e) < \min(|N_{0}(e)|, q_0(e)) + |\mu_1(e)|$. Additionally, we know that we set $q_1'(e) = q_0(e) - \min(|N_0(e)|,q_0(e))$, and since $|\mu_1(e)| \leq q_1'(e)$, we finally get that $q_0(e) \leq q_0(e)$, which is a contradiction.

Therefore, by contradiction, $\mu_0$ is a maximal matching on $G_0$.
\end{proof}

Given this, we now show that the second matching $\mu_1$ could be equivalent between \prm{} and \sprm{}.

\begin{lem}\label{lem:smp1}

Let $G_0 = (V_0, E_0, q_0)$ and $G_1 = (V_1, E_1, q_1)$ be the first  and second subgraphs considered by \textsf{PriorityMatch} with capacities $q_0$ and $q_1$. Assume an implementation of \prm{} and \sprm{} result in the same matching on $G_0$, call it $\mu_0$. If $\mu_1$ is the matching found by \textsf{SmartPriorityMatch} on $G_1$, then $\mu_1$ is a maximal matching on $G_1$.
\end{lem}

\begin{proof}
Consider the same notation as introduced in Lemma~\ref{lem:smp0} and fix some $e\in E$. Recall in Lemma~\ref{lem:smp0} we showed that $|\mu_0(e)| = \min(|N_0(e)|, q_0(e))$. Therefore, $e$'s quota in $G_1$ for \prm{} is $q_1(e) = q_0(e) - |\mu_0(e)| = q_0(e) - \min(|N_0(e)|, q_0(e))$. In \sprm{}, its quota is $q_1'(e) = q_0 - \min(|N_0(e)|, q_0(e))$. Therefore, for all $e\in E$, $q_1(e) = q_1'(e)$.

Now we will simply show that it is both a valid matching and it is maximal in $G_1$. Assume for contradiction it is not a valid matching in $G_1$. That means it must match some $a\in A$ (as we know all $e\in E$ has the same capacity in both graphs) above its capacity. Note that $q_{1}'(a)=q_{0}(a)$, and $q_{1}(a) \geq q_{1}'(a)-1$, because since $a$ could only be the leaf of a star component in $G_0$, $|\mu_0(a)| \leq 1$. Therefore, $a$'s capacity could only have been exceeded by 1, and this only occurs when $|\mu_0(a)| = 1$ and \textsf{SmartPriorityMatch} constructs $\mu_1$ such that $|\mu_1(a)| = q_{0}'(a)$. For this to happen, then, \textsf{SmartPriorityMatch} first matches $a$ up to its capacity in $\mu_1$. Then $a$'s capacity is reduced to 0, so it cannot match $a$ in $\mu_0$. This contradicts that $|\mu_0(a)| =1$. Thus, $\mu_1$ is a valid matching on $G_1$.

Assume for contradiction that $\mu_1$ is not maximal. This implies there is some $(e,a)\in G_1\setminus\mu_1$ where both $e$ and $a$ are not matched up to their capacity in $G_1$. Note, however, since $\mu_1$ is maximal according to the reservation, either $e$ is matched to capacity, $a$ is matched to capacity, or $a$ is reserved. If $e$ is matched to capacity in the reserved matching, then it must also be matched to capacity in $G_1$ since its capacity is the same in both. This is a contradiction. Otherwise, $a$ is matched to capacity in the reserved matching or it is reserved. If $a$ is matched to capacity, notice that since $q_1'(a) = q_0(a)$, thus $|\mu_1(a)| = q_0(a) \geq q_1(a)$. Thus, $a$ must be (at least) matched to capacity in $G_1$. Finally, we consider when $a$ is reserved. Let $e = \aff^{-1}(a)$. For $a$ to have remaining capacity and a remaining unmatched neighbor with capacity and be forced to not match, it could have only had 1 remaining capacity ($q_1'(a) - |\mu_1(a)| = 1$) and there must have been exactly $r(N_0(e)) = \min(|N_0(e)|, q_0(e))$ agents in $N_0(e)$ that were not matched to capacity by $\mu_1$. Since $|\mu_0(e)| = \min(|N_0(e)|, q_0(e))$ and all other $a'\in N_0(e)$ must have had $q_0'(a') = 0$, these agents must precisely make up $\mu_0(e)$. Thus, $|\mu_0(a)| = 1$. Therefore, $|\mu_0(a)| + |\mu_1(a)| = 1 + q_1'(a) - 1 = q_1'(a) = q_0(a)$. Thus, in \prm{}, $a$ has met its capacity in $G_1$. This is a contradiction

Thus by contradiction, $\mu_1$ is maximal on $G_1$.
\end{proof}

Now we can immediately prove Lemma~\ref{lem:smp}.

\begin{proof}[Proof of Lemma~\ref{lem:smp}]
Simply combine Lemmas~\ref{lem:smp0} and \ref{lem:smp1}, and note that after $\mu_0$ and $\mu_1$ are found, \sprm{} acts identically to \prm{}. This is sufficient to show \textsf{SmartPriorityMatch} is a valid implementation of \textsf{PriorityMatch}.
\end{proof}

Now we must prove \textsc{SmartPriorityMatch} exhibits additional properties to \textsc{PriorityMatch}. We start by showing it satisfies the preconditions for Lemma~\ref{lem:prior}.

\begin{lem}\label{lem:precond}
Let $\mu$ be the matching obtained by \textsf{SmartPriorityMatch}. Consider a potential blocking tuple $\mathcal{T} = (a,a',a'',e,e',e'')$ such that $a,a'\in \aff(e)$. Let $\mu'$ be the swapped matching of $\mu$ with respect to $\mathcal{T}$. Then if $a$ prefers $\mu'$ to $\mu$:
\[\pr_e^e(a') + \lambda\pr_e^a(e') + \lambda\pr_e^{a'}(e) \geq \pr_e^e(a) + \lambda\pr_e^a(e) + \lambda\pr_e^{a'}(e'').\]
\end{lem}

\begin{proof}
For the beginning of the proof, we will be viewing the order of events with respect to \prm{}. To that end, $\mu_0$ was formed first, then $\mu_1$ and the rest.

Fix our tuple and matchings $\mu$ and $\mu'$ and assume $a,a'\in\aff(e)$ and $a$ prefers $\mu'$ to $\mu$. Since $a$ prefers $\mu'$ to $\mu$, we know by Lemma~\ref{lem:prefs} that $\pr_a(e) > \pr_a(e')$, which means $\pr_a(e) = 1$ and $\pr_a(e') = 0$. Since $a$ cannot match to something it does not like in \sprm{}, it must be the case that $e'\in\mathcal{E}$ (and thus $a''\in\mathcal{E}$), and so $a$ never matched to its quota.

Since $a'\in \aff(e)$, it must exist in the tuple. By the potential blocking tuple definition, $(a',e)\in \mu$. Thus it must exist in some graph. Since $a'\in \aff(e)$, it must have been $G_0$, $G_2$, or $G_3$. If $(e,a)$ existed in a graph $G_i$, since they did not match but $a$ never reached its capacity, that means $e$ must have reached its capacity in or before $\mu_i$ Thus, $(a',e)$ must have appeared at latest in graph $G_i$. It is not hard to see since $\lambda \leq 1$ that this implies that:

\[\pr_e^e(a') + \lambda\pr_e^{a'}(e) \geq \pr_e^e(a) + \lambda\pr_e^a(e)\]

Thus, to prove the lemma, it is sufficient to show that $\pr_e^{a'}(e'') = 0$. Assume for contradiction it is 1. For this to be true, $e''\notin\mathcal{E}$. Additionally, by the definition of a blocking tuple, $(a',e'')\notin \mu$, $\pr_{a'}(e'') = 1$, and (since $a'\notin\aff(e'')$ because it is in $\aff(e)$) $\pr_{e''}^{e''}(a') = 1$. Therefore, $(a',e'')$ appeared in $G_1$. For them to not match, that means $a'$ must not have quota after $G_1$, so all its matches must have been in $G_0$ and $G_1$. Therefore $(a',e)$ must have occurred in $G_0$.

We will now view matchings in the order that occurs in \sprm{}, so $\mu_1$ occurs first, then $\mu_0$ and the rest. Recall from Lemma~\ref{lem:smp0} that $|\mu_0(e)| = \min(|N_0(e)|,q_0(e)) = r(N_0(e))$. Therefore, the set of agents $\mu_0(e)$ must have not been matched to quota when $\mu_1$ was made (before $\mu_0$ was made). Thus, $\mu_0(e)$ is sufficient to satisfy the reservation on $N_0(e)$  for the matching $\mu_1(e)$. Since $(a,e)$ was not matched, $a\notin\mu_0(e)$ even though $(a,e)\in G_0$, meaning $a\in N_0(e)$. Additionally, $a$ was never matched to capacity. Therefore, there are at least $r(N_0(e)) + 1$ agents in $N_0(e)$ that did not meet their capacity in $\mu_1$. Additionally, since $|\mu_0(a')| = 1$, $|\mu_1(a')| \leq q_0(a') - 1$. Since $\mu_1$ occurred first and the reservation constraint had not been met and $e''$ had capacity (since it never matched to capacity), $(a',e'')$ would have matched in $\mu_1$. This is a contradiction. This concludes our proof.
\end{proof}

Now we proceed with Theorem~\ref{thm:match}.

\begin{proof}[Proof of Theorem~\ref{thm:match}]
Lemma~\ref{lem:smp} shows that \textsf{SmartPriorityMatch} is a specific implementation of \textsf{PriorityMatch}. Additionally, Lemmas~\ref{lem:precond} shows that \textsf{SmartPriorityMatch} satisfies the preconditions of Lemma~\ref{lem:prior}. Therefore, by Lemma~\ref{lem:prior}, \textsf{SmartPriorityMatch} finds a stable matching.
\end{proof}

%% file: spm_pseudocode.tex
\subsection{\textsf{SmartPriorityMatch} Pseudocode}\label{app:pseudocode}

Here we present the pseudocode for \textsf{SmartPriorityMatch}. This can be seen in Algorithm~\ref{alg:spm}.

\begin{algorithm}
 \renewcommand{\algorithmicrequire}{\textbf{Input: }}
	\renewcommand{\algorithmicensure}{\textbf{Output: }}
    \caption{\textsf{SmartPriorityMatch}}
    \label{alg:spm}
    \begin{algorithmic}[1]
    	\REQUIRE Sets $A$ and $E$ of agents, affiliate function $\aff: e \to \mathcal{P}(A)$, quota function $q:A\cup E\to \mathbb{N}$, preference functions $\forall a\in A\,\, \pr_a:E\to\{0,1\}$, $\forall e\in E\,\, \pr_e^e:A\to\{0,1\}$, $\forall e\in E, a\in \aff(e)\,\, \pr_e^a:E\to\{0,1\}$, and a string $val$ to designate the valuation function
        \ENSURE A stable matching $\mu$
        \STATE $V\gets E\cup A$
        \STATE $E_0 \leftarrow \{ (a,e)\in A\times E: a\in\aff(e),\pr_a(e) = \pr_e^e(a) = \pr_e^a(e) = 1\}$
        \STATE $\forall e\in E, N_0(e) = \{a\in A: (e,a)\in E_0\}$
        \STATE $E_1 \gets \{(a,e)\in A\times E: a\notin\aff(e), \pr_a(e)=\pr_e^e(a)=1\}$
        \STATE $\forall e\in E, q_1'(e) \gets q(e) - \min(|N_0(e)|, q(e))$
        \STATE $\forall a\in A, q_1'(a) \gets q(a)$
        \STATE $\mathcal{S} = \{N_{0}(e): e\in E\}$
        \STATE $\forall e\in E\,\, r(N_{0}(e)) = \min(|N_{0}(e)|, q(e))$
        \STATE $\mu_1\gets \textsf{ReservedMaximalMatching}(V, E_1, q_1', \mathcal{S}, r)$
        \STATE $\forall a\in E\cup A, q_0'(a) \gets q(a) - |\mu_1(a)|$
        \STATE $\mu_0 \gets \textsf{MaximalMatching}(V, E_0, q_0')$
        \STATE $E_2 \gets \{(a,e)\in A\times E: \pr_a(e) = 1, \pr_e^e(a) \neq \pr_e^a(e)\}$
        \STATE $\forall a\in E\cup A, q_2(a) \gets q(a) - |\mu_0(a)| - |\mu_1(a)|$
        \STATE $\mu_2 \gets \textsf{MaximalMatching}(A\cup E, E_2, q_2')$
        \STATE $E_3 \gets \{(a,e)\in A\times E: \pr_a(e) = \pr_e^a(e) = 1, \pr_e^e(a) =0\}$
        \STATE $\forall a\in E\cup A, q_3(a) \gets q_2(a) - |\mu_2(a)|$
        \STATE $\mu_3 \gets \textsf{MaximalMatching}(A\cup E, E_3, q_3')$
        \RETURN $\mu_0\cup\mu_1\cup\mu_2\cup\mu_3$.
    \end{algorithmic}
\end{algorithm}

%% file: ilp_proofs.tex
\section{ILP Formulation and Proofs (\S\ref{sec:experiments})}\label{app:ilp}

In this section, we formulate our problem as an integer linear program (ILP). As this is a rather standard and straightforward solution, and ILP solvers are known to be efficient in practice, this is a good baseline to compare our algorithm to. Let $z_{e,a}$ for all $e\in E$ and $a\in A$ denote $(e,a)$ is matched if $z_{e,a} = 1$ and $(e,a)$ is unmatched if $z_{e,a} = 0$. Our basic constraints are as follows:
\begin{align*}
\forall e\in E, a\in A :& z_{e,a} \in \{0,1\}\tag{1}
\\\forall e\in E :& \sum_{a\in A} z_{e,a} \leq q(e)\tag{2}
\\\forall a\in A :& \sum_{e\in E} z_{e,a} \leq q(a)\tag{3}
\end{align*}
These constraints simply ensure all edges are matched or not and the number of matches containing an agent does not exceed that agent's capacity. Note that this is sufficient to ensure that we have a valid matching. Now we need to consider stability. To do this concisely, we introduce a function $\textsf{coeff}:\mathcal{B}\to \mathbb{N}$, where $\mathcal{B}$ is the set of tuples $\mathcal{T} = (a,a',a'',e,e',e'')$ that satisfy all conditions for being a blocking tuple except those that depend on $\mu$ (i.e., if we introduce the appropriate $\mu$, $\mathcal{T}$ is a blocking tuple). Note that our definition of $\mathcal{B}$ determines the weight selection for our valuation function, as it determines which potential blocking tuples could actually be blocking tuples based off of preferences. In this program, we will use many indicators. To refrain from introducing many new variables, we let $\mathbb{I}_{p}$ for some boolean $p$ be 1 if $p$ is true, and 0 otherwise. For example, $\mathbb{I}_{a',e'\in\mathcal{E}}$ is 1 if $a'$ and $e'$ are both in $\mathcal{E}$, and otherwise it is 0. Then we can define \textsf{coeff} as follows:
\begin{align*}
\textsf{coeff}&(a,a',a'',e,e',e'') \\=& q(e)q(a)\mathbb{I}_{a',e'\in\mathcal{E}} + q(e)\mathbb{I}_{a'\in\mathcal{E},e'\notin\mathcal{E}} 
+ q(a)\mathbb{I}_{a'\notin\mathcal{E},e'\in\mathcal{E}} 
\\&+ \mathbb{I}_{a',e'\notin\mathcal{E}, a'',e''\in\mathcal{E}} + q(e'')q(a'') \mathbb{I}_{a'',e''\notin \mathcal{E}\cup \{a',e'\}}
\\&+ q(e'')\mathbb{I}_{e''\notin \mathcal{E}, a''\in\mathcal{E}}
+ q(a'')\mathbb{I}_{e''\in \mathcal{E}, a''\notin\mathcal{E}}
\end{align*}
This can also be thought of a conditional, where we return $q(e)q(a)$, $q(e)$, $q(a)$, 1, $q(e'')q(a'')$, $q(e'')$, or $q(a'')$ depending on which elements in the tuple are in $\mathcal{E}$ or not. Note that these are all constants: it does not involve variables from the ILP. Then our constraints to ensure stability are as follows, where:
\begin{align*}
\forall& \mathcal{T}=(a,a',a'',e,e',e'')\in \mathcal{B}:\\
&\textsf{coeff}(\mathcal{T})z_{e,a} + \mathbb{I}_{a'\notin\mathcal{E}}\textsf{coeff}(\mathcal{T})(1-z_{e,a'}) 
\\&+ \mathbb{I}_{e'\notin\mathcal{E}}\textsf{coeff}(\mathcal{T})(1-z_{e',a}) +\mathbb{I}_{a''\notin\mathcal{E}}\textsf{coeff}(\mathcal{T})z_{e',a''} 
\\&+ \I_{e''\notin\E}\textsf{coeff}(\mathcal{T})z_{e'',a'} + \frac{\textsf{coeff}(\mathcal{T})}{q(a)} \I_{e'\in\E} \sum_{e^*\in E} z_{e^*,a} 
\\&+ \frac{\textsf{coeff}(\mathcal{T})}{q(e)} \I_{a'\in\E} \sum_{a^*\in A} z_{e,a^*}
+ \frac{\textsf{coeff}(\mathcal{T})}{q(a'')} \I_{a''\notin\E} \sum_{e^*\in E} z_{e^*,a''} 
\\&+ \frac{\textsf{coeff}(\mathcal{T})}{q(e'')} \I_{e''\notin\E} \sum_{a^*\in A} z_{e'',a^*}
\\&\qquad\qquad\qquad\qquad\qquad\qquad\qquad\qquad\leq \textsf{coeff}(\T)\tag{4}
\end{align*}
While these constraints may seem construed, they are derived directly from the definition of a blocking tuple, and account for all possible cases of a blocking tuple. Therefore, this is the most direct translation of the \bamprob{} into an ILP.

\begin{theorem}\label{thm:ilp}
The ILP defined by (1), (2), (3), and (4) solves the \bamprob{}.
\end{theorem}

We prove Theorem~\ref{thm:ilp} by breaking down the construction of the ILP.

\begin{proof}[Proof of Theorem~\ref{thm:ilp}]
We start by constructing the ILP from the ground up, and it will be easy to check (albeit, time-consuming) that this ILP is just a broken down version of the ILP in question. Let $z_{e,a}$ for all $e\in E$ and $a\in A$ denote $(e,a)$ is matched if $z_{e,a} = 1$ and $(e,a)$ is unmatched if $z_{e,a} = 0$. Our basic constraints are as follows:

\begin{align*}
\forall e\in E, a\in A :& z_{e,a} \in \{0,1\}
\\\forall e\in E :& \sum_{a\in A} z_{e,a} \leq q(e)
\\\forall a\in A :& \sum_{e\in E} z_{e,a} \leq q(a)
\end{align*}

These constraints simply ensure all edges are matched or not, and the number of matches containing an agent does not exceed its capacity. Note that this is sufficient to ensure that we have a valid matching. Next, we must ensure stability. We will consider a number of potential blocking tuples. We break it down into all the different ways matches can be broken down and reformed for a swapped matching.

First: consider when some $a\in A$ and $e\in E$ are simply undermatched. Then, without breaking matches, they are allowed to match to each other. They will only do this if they get something out of the match. We must ensure that either they are matched together, or one has reached capacity. For notation, let $I_a^e$ be 1 if $a\in\aff(e)$ and 0 otherwise. We can guarantee the result with the following constraint:

\begin{align*}
\forall &a\in A, e\in E \text{ s.t. } \pr_a(e)  = 1 \land \pr_e^e(a) + I_a^e\pr_e^a(e) \geq 1: \\&(q(a)\cdot q(e))z_{e,a} + q(a)\sum_{a^*\in A} z_{e,a^*} + q(e)\sum_{e^*\in E} z_{e^*,a} \\&\qquad\qquad\qquad\qquad\qquad\qquad\qquad\qquad\geq q(a) \cdot q(e)
\end{align*}

Second: consider when $e$ is undermatched, but $a$ drops its match with some $e'\in\mu(a)$. These three together will only form a blocking tuple if $a$ prefers $e$ to $e'$ and $e$ gets something out of the match. We must ensure that $(a,e)$ is matched or $(a,e')$ is not matched or $e$ is at capacity:

\begin{align*}
&\forall a\in A, e\in E, e'\in E\setminus\{e\} \\&\qquad\text{ s.t. } \pr_a(e) > \pr_a(e') \land \pr_e^e(a)  + I_a^e\pr_e^a(e) \geq I_{a}^e\pr_e^{a}(e'): 
\\&q(e)z_{e,a} + q(e)(1-z_{e',a}) + \sum_{a^*\in A} z_{e,a^*} \geq q(e)
\end{align*}

Third: consider when $a$ is undermatched, but $e$ drops its match with some $a'\in \mu(e)$. These three together will only form a blocking tuple if $e$ prefers a match from $a$ to $e$ than $a'$ to $e$ and $a$ gets something out of the match. We must ensure that $(a,e)$ is matched or $(a',e)$ is not matched or $a$ is at capacity:

\begin{align*}
&\forall a\in A, e\in E, a'\in A\setminus\{a\} \\&\qquad\text{ s.t. } \pr_a(e) = 1 \\&\qquad\qquad\land \pr_e^e(a)  + I_a^e\pr_e^a(e) \geq \pr_e^e(a') + I_{a'}^e\pr_e^{a'}(e): 
\\&q(a)z_{e,a} + q(a)(1-z_{e,a'}) + \sum_{e^*\in E} z_{e^*,a} \geq q(a)
\end{align*}

Fourth: consider when $a$ and $e$ drop matches $e'$ and $a'$ respectively to match with each other, but neither $e'$ nor $a'$ decide to rematch. This is only notable when both $a$ and $e$ prefer being matched together. In this case, we must ensure $(a,e)$ is matched or $(a,e')$ is not matched or $(a',e)$ is not matched.

\begin{align*}
&\forall a\in A, e\in E, a'\in A\setminus\{a\}, e'\in E\setminus\{e\} \\&\,\,\text{ s.t. } \pr_a(e) > \pr_a(e') \\&\quad\land \pr_e^e(a)  + I_a^e\pr_e^a(e) \geq \pr_e^e(a') + I_{a}^e\pr_e^{a}(e') + I_{a'}^e\pr_e^{a'}(e): 
\\&z_{e,a} + (1-z_{e,a'}) + (1-z_{e',a}) \geq 1
\end{align*}

Fifth: consider when $a$ and $e$ drop matches $e'$ and $a'$ respectively to match with each other, and $a'$ rematches with some $e''$ that is undermatched and $e'$ does not rematch. This is only notable when both $a$ and $e$ prefer being matched together and both $a'$ and $e''$ gain from being matched together. We need to ensure that $(a,e)$ is matched, $(a,e')$ is not matched, $(a',e)$ is not matched, $(a',e'')$ is matched, or $e''$ is matched to capacity:

\begin{align*}
&\forall a\in A, e\in E, a'\in A\setminus\{a\}, e'\in E\setminus\{e\}, e''\in E\setminus\{e,e'\} \\&\,\,\text{ s.t. } \pr_a(e) > \pr_a(e') \\&\quad\land \pr_e^e(a)  + I_a^e\pr_e^a(e) + I_{a'}^e\pr_e^{a'}(e'')\\&\qquad\qquad\geq \pr_e^e(a') +  I_{a}^e\pr_e^{a}(e') + I_{a'}^e\pr_e^{a'}(e)
\\&\quad \land \pr_{a'}(e'') = 1 \land \pr_{e''}^{e''}(a') + I_{a'}^{e''} \pr_{e''}^{a'}(e'') \geq 1: 
\\&q(e'')z_{e,a} + q(e'')(1-z_{e,a'}) + q(e'')(1-z_{e',a})  \\&\qquad\qquad\qquad\qquad+ q(e'')z_{e'',a'} + \sum_{a^*\in A} z_{e'',a^*} \geq q(e'')
\end{align*}

Sixth: consider when $a$ and $e$ drop matches $e'$ and $a'$ respectively to match with each other, and $e'$ rematches with some $a''$ that is undermatched and $a'$ does not rematch. This is only notable when both $a$ and $e$ prefer being matched together and both $e'$ and $a''$ gain from being matched together. We need to ensure that $(a,e)$ is matched, $(a,e')$ is not matched, $(a',e)$ is not matched, $(a'',e')$ is matched, or $a''$ is matched to capacity:

\begin{align*}
&\forall a\in A, e\in E, a'\in A\setminus\{a\}, e'\in E\setminus\{e\}, a''\in A\setminus\{a,a'\} \\&\,\,\text{ s.t. } \pr_a(e) > \pr_a(e') \\&\quad\land \pr_e^e(a)  + I_a^e\pr_e^a(e) + I_{a''}^e\pr_e^{a''}(e')\\&\qquad\qquad\geq \pr_e^e(a') + I_{a}^e\pr_e^{a}(e') + I_{a'}^e\pr_e^{a'}(e)
\\&\quad \land \pr_{a''}(e') = 1 \land \pr_{e'}^{e'}(a'') + I_{a''}^{e'} \pr_{e'}^{a''}(e') \geq 1: 
\\&q(a'')z_{e,a} + q(a'')(1-z_{e,a'}) + q(a'')(1-z_{e',a})  \\&\qquad\qquad\qquad\qquad+ q(a'')z_{e',a''} + \sum_{e^*\in E} z_{e^*,a''} \geq q(a'')
\end{align*}

Seventh: consider when $a$ and $e$ drop matches $e'$ and $a'$ respectively to match with each other, and $a'$ and $e'$ rematch with some $e''$ and $a''$ respectively that are either both undermatched or $a''=a'$ and $e''=e'$. This is only notable when both $a$ and $e$ prefer being matched together, both $a'$ and $e''$ gain from being matched together, and both $e'$ and $a''$ gain from being matched together. We need to ensure that $(a,e)$ is matched, $(a,e')$ is not matched, $(a',e)$ is not matched, $(a',e'')$ is matched, $(a'', e')$ is matched, or if $a''\neq a'$ and $e''\neq e'$, then either $a''$ or $e''$ is matched to capacity (recall $J_{a'',a'} = 1$ if and only if $a''\neq a'$):

\begin{align*}
&\forall a\in A, e\in E, a'\in A\setminus\{a\}, e'\in E\setminus\{e\}, a''\in A\setminus\{a,a'\} \\&\,\,\text{ s.t. } \pr_a(e) > \pr_a(e') \\&\quad\land \pr_e^e(a)  + I_a^e\pr_e^a(e) + I_{a''}^e\pr_e^{a''}(e')\geq I_{a}^e\pr_e^{a}(e') + I_{a'}^e\pr_e^{a'}(e)
\\&\quad \land \pr_{a'}(e'') = 1 \land \pr_{e''}^{e''}(a') + I_{a'}^{e''} \pr_{e''}^{a'}(e'') \geq 1
\\&\quad \land \pr_{a''}(e') = 1 \land \pr_{e'}^{e'}(a'') + I_{a''}^{e'} \pr_{e'}^{a''}(e') \geq 1: 
\\&(q(a'')\cdot q(e''))z_{e,a} + (q(a'')\cdot q(e'')) (1-z_{e,a'}) \\&+ (q(a'') \cdot q(e''))(1-z_{e',a})  + (q(a'')\cdot q(e''))z_{e',a''} \\&+(q(a'')\cdot q(e''))z_{e'',a'} \\&+ J_{a'',a'}\left(q(e'')\sum_{e^*\in E} z_{e^*,a''} + q(a'')\sum_{a^*\in A} z_{e'',a^*}\right) \\&\qquad\qquad\qquad\geq (q(a'')\cdot q(e''))
\end{align*}

This encodes all the cases for the existence of a blocking tuple. Therefore, solutions to the ILP directly correspond to solutions to the dichotmous affiliate stable matching problem. Note that the ILP of interest is actually equivalent to this, one simply needs to go through each type of blocking tuple and check the inequalities.
\end{proof}